\documentclass[twocolumn,superscriptaddress,amsmath,amssymb,nofootinbib]{revtex4-1}

\pdfoutput=1
\usepackage{graphicx}% Include figure files
\usepackage{xcolor}
\usepackage{xspace} % for abbreviations
\usepackage{epstopdf}
\usepackage[mathscr]{euscript}
\usepackage{mathtools} % get multilined command
\usepackage{textcomp} % \textdegree command
\usepackage{dcolumn}% Align table columns on decimal point
\usepackage{bigstrut}
\usepackage{float}
\usepackage[english]{babel}
\usepackage{bm}% bold math
\usepackage[mathlines]{lineno}
\usepackage{multirow}
\usepackage{lipsum}  
\usepackage[section] {placeins} 
\usepackage{glossaries}
\usepackage{tabularx} 
\usepackage{hyperref}
\hypersetup{colorlinks,allcolors=black}
\begin{document}
\newcommand*{\ddoi}[1]{\href{http://dx.doi.org/#1}{\texttt{doi: #1}}}
\newcommand*{\uurl}[1]{\href{#1}{\texttt{url: #1}}}
\newcommand*{\purl}[2]{\href{#1}{\texttt{#2}}}
\newcommand{\element}[2]{{$^{\rm #1}{\rm #2}$}}

\newcommand{\deltachisq}{$\Delta\chi^2$}
\newcommand{\chisq}{$\chi^2$}
\newcommand{\kevt}{keV$_{\rm t}$}
\newcommand{\kevee}{keV$_{\rm ee}$}
\newcommand{\evee}{eV$_{\rm ee}$}
\newcommand{\eh}{$e^-h^+$}
\newcommand{\gevc}{GeV/${\rm c}^{\rm 2}$}
\newcommand{\degr}{$^\circ$}

\newacronym{adc}{ADC}{analog-to-digital converter}
\newacronym{auc}{AUC}{area under the curve}
\newacronym[\glslongpluralkey={Class Activation Maps}, \glsshortpluralkey={CAMs}]{cam}{CAM}{Class Activation Map}
\newacronym{cdms}{CDMS}{Cryogenic Dark Matter Search}
\newacronym{cdmslite}{\mbox{CDMSlite}}{Cryogenic Dark Matter Search Low Ionization Threshold Experiment}
\newacronym{cmb}{CMB}{Cosmic Microwave Background}
\newacronym[\glslongpluralkey={Convolutional Neural Networks}, \glsshortpluralkey={CNNs}]{cnn}{CNN}{Convolutional Neural Network}
\newacronym{daq}{DAQ}{data acquisition}
\newacronym{dm}{DM}{dark matter}
\newacronym{er}{ER}{electron recoil}
\newacronym{fpr}{FPR}{false positive rate}
\newacronym{gap}{GAP}{global average pooling}
\newacronym{hv}{HV}{high voltage}
\newacronym{izip}{iZIP}{interleaved Z-Sensitive Ionization and Phonon}
\newacronym{lfn}{LFN}{low-frequency noise}
\newacronym[\glslongpluralkey={Long Short-Term Memorys}, \glsshortpluralkey={LSTMs}]{lstm}{LSTM}{Long Short-Term Memory}
\newacronym{cnnlstm}{CNNLSTM}{Convolutional Neural Network with Long Short-Term Memory}
\newacronym{mse}{MSE}{mean square error}
\newacronym{ntl}{NTL}{Neganov-Trofimov-Luke}
\newacronym{nsof}{NSOF}{Non-Stationary Optimal Filter}
\newacronym{nr}{NR}{nuclear recoils}
\newacronym{oi}{OI}{optimum interval}
\newacronym{of}{OF}{optimal filter}
\newacronym{pbs}{PBHs}{primordial black holes}
\newacronym{plr}{PLR}{profile likelihood ratio}
\newacronym[\glslongpluralkey={power spectral densities}, \glsshortpluralkey={PSDs}]{psd}{PSD}{power spectral density}
\newacronym{pt}{PT}{phonon total}
\newacronym[\glslongpluralkey={Quasiparticle-Trap-Assisted Electrothermal-Feedback Transition Edge Sensors}, \glsshortpluralkey={QETs}]{qet}{QET}{Quasiparticle-Trap-Assisted Electrothermal-Feedback Transition Edge Sensor}
\newacronym{rnlt}{RNTLs}{Reduced NTL-Gain Events}
\newacronym{roc}{ROC}{receiver operating characteristic}
\newacronym{sd}{SD}{spin-dependent}
\newacronym{si}{SI}{spin-independent}
\newacronym{supr}{SUPR}{Simulation Using Pulse-shapes and Regression}
\newacronym[\glslongpluralkey={Transition Edge Sensors}, \glsshortpluralkey={TESs}]{tes}{TES}{transition edge sensor}
\newacronym{tpr}{TPR}{true positive rate}
\newacronym[\glslongpluralkey={Weakly Interacting Massive Particles}, \glsshortpluralkey={WIMPs}]{wimp}{WIMP}{Weakly Interacting Massive Particle}
\newacronym{roi}{ROI}{region of interest}
\newacronym{ml}{ML}{machine learning}
\newacronym{tp}{TP}{true positives}
\newacronym{fp}{FP}{false positives}
\newacronym{tn}{TN}{true negatives}
\newacronym{fn}{FN}{false negatives}
\title{Low Frequency Noise Identification for the SuperCDMS Experiment\texorpdfstring{\\}{ }using Convolutional Neural Networks}
\author{M.F.~Albakry} \affiliation{Department of Physics \& Astronomy, University of British Columbia, Vancouver, BC V6T 1Z1, Canada}\affiliation{TRIUMF, Vancouver, BC V6T 2A3, Canada}
\author{I.~Alkhatib} \affiliation{Department of Physics, University of Toronto, Toronto, ON M5S 1A7, Canada}
\author{D.~Alonso-González} \affiliation{Departamento de F\'{\i}sica Te\'orica, Universidad Aut\'onoma de Madrid, 28049 Madrid, Spain}\affiliation{Instituto de F\'{\i}sica Te\'orica UAM-CSIC, Campus de Cantoblanco, 28049 Madrid, Spain}
\author{J.~Anczarski} \affiliation{SLAC National Accelerator Laboratory/Kavli Institute for Particle Astrophysics and Cosmology, Menlo Park, CA 94025, USA}
\author{T.~Aralis} \affiliation{SLAC National Accelerator Laboratory/Kavli Institute for Particle Astrophysics and Cosmology, Menlo Park, CA 94025, USA}
\author{T.~Aramaki} \affiliation{Department of Physics, Northeastern University, 360 Huntington Avenue, Boston, MA 02115, USA}
\author{A.~Ashtari~Esfahani} \affiliation{TRIUMF, Vancouver, BC V6T 2A3, Canada}
\author{I.~Ataee~Langroudy} \affiliation{Department of Physics and Astronomy, and the Mitchell Institute for Fundamental Physics and Astronomy, Texas A\&M University, College Station, TX 77843, USA}
\author{R.~Bhattacharyya} \affiliation{Department of Physics and Astronomy, and the Mitchell Institute for Fundamental Physics and Astronomy, Texas A\&M University, College Station, TX 77843, USA}
\author{P.L.~Brink} \affiliation{SLAC National Accelerator Laboratory/Kavli Institute for Particle Astrophysics and Cosmology, Menlo Park, CA 94025, USA}
% ...

\smallskip
\date{\today}
\collaboration{SuperCDMS Collaboration}
\noaffiliation
\smallskip
\begin{abstract}
\vspace{0.6em}
The SuperCDMS experiment uses semiconductor crystal detectors operated at cryogenic temperatures to search for low-mass dark matter. Vibrations observed during the SuperCDMS Soudan experiment generated broadband \acrfull{lfn}, which due to its similarity in the pulse shape to the low-energy signal events is difficult to remove. In the final low ionization threshold analysis, CDMSlite Run\,3, the strong event selection criterion applied to remove LFN events also significantly reduced the ability to identify signal events in the low energy region, thus reducing the sensitivity of the experiment to low-mass dark matter. Therefore, an LFN rejection criterion using machine learning (ML) aimed to improve the signal efficiency while maintaining the signal purity has now been developed. This paper presents a new method using \acrfullpl{cnn}, its application to the CDMSlite Run\,3 dataset, and its impact on dark matter sensitivities, with further investigation through case studies. 
\end{abstract}

\maketitle
\section{SuperCDMS experiment at Soudan}

The SuperCDMS Soudan experiment \cite{soudan_1, soudan_2} used germanium (Ge) crystal detectors cooled to $\sim$50\,mK temperature to search for \acrfull{dm} particle interactions with Ge nuclei. The experiment was conducted in the Soudan Mine in Minnesota, USA, between 2009 and 2015. The detectors, which are configured to incorporate both phonon and charge readout channels, are known as \acrfull{izip} detectors. 

Particle interactions can scatter off nuclei and produce phonons at the interaction site, referred to as ``prompt phonons". These interactions can also ionize atoms and create electron–hole (eh) pairs. When a bias voltage is applied across the detector, these charges travel through the crystal under the effect of the electric field. The drifting eh pairs generate additional phonons, thereby amplifying the phonon signal via the \acrfull{ntl} effect  \cite{ntl1,ntl2}. The design voltage of the \acrshort{izip} detectors is only a few volts, so the amplification from the NTL effect is comparable to the recoil energy. The phonon sensors, located on the top and bottom surfaces of the detector, are Quasiparticle-trap-assisted Electrothermal-feedback Transition-edge Sensors (QETs) \cite{qet}. They consist of superconducting aluminum fins that absorb athermal phonons and the absorbed energy is then measured as heat by the \acrfullpl{tes} \cite{tes}. 

\subsection{ Cryogenic Dark Matter Search Low Ionization Threshold Experiment (CDMSlite) }

\begin{figure}
	\centering
	\includegraphics[width=0.6\linewidth]{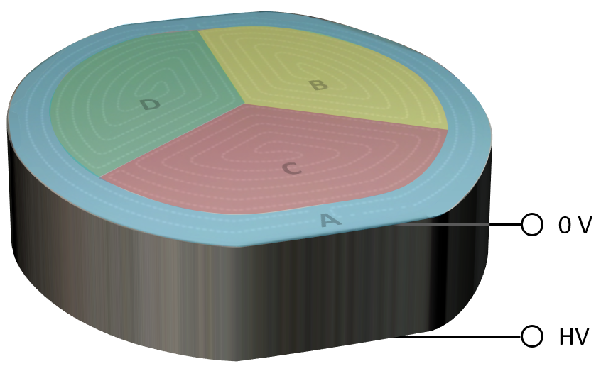}
	\caption{Schematic of the CDMSlite detector showing the layout and labeling of the phonon channels (A, B, C, and D). In this configuration, the one side of the crystal was applied a high voltage (HV) bias and the other side was held at ground. Only the phonon channels on the grounded side were read out. \cite{cdmsliter2}}
	\label{fig:detector}
\end{figure}

The SuperCDMS Soudan experiment consisted of 15 germanium iZIP detectors. In the final phase of the experiment, the detectors were operated at a voltage higher than its design value up to 80\,V in order to strongly leverage the \acrshort{ntl} effect. As the \acrshort{ntl} amplification scales linearly with the applied bias voltage, small energy deposits can be significantly amplified. This mode of detector operation is referred to as the \acrfull{cdmslite} \cite{cdmslite_1, cdmslite_2,cdmsliter2,CDMSliter3}. The phonon sensors are grouped into four readout channels labeled as PA, PB, PC, and PD. The charge channels in the CDMSlite configuration were only used for removing events with significant noise in the waveforms and were not used for event reconstruction. The schematics of the detector and its phonon channel layout are shown in Fig.~\ref{fig:detector}. The final  data-taking period (run) of the experiment, CDMSlite Run\,3, conducted between February and May 2015, where one of the detectors was operated at $\sim$75\,V, is used for the analysis presented in this publication. The CDMSlite Run\,3 experiment had an  exposure of 60.9 days, which corresponds to 36.4\,$\pm$\,0.3\,kg-days with a detector mass of 0.597\,$\pm$\,0.005\,kg.

The cryocooler, which was installed to minimize the heat load from the warm electronics going into the cryostat, degraded over the course of the run.
This introduced vibrations that were transmitted to the detector via the electronics stem, disturbing the baseline of the waveforms and resulting in higher noise in the low-frequency region. The associated baseline fluctuations frequently triggered the detector, contaminating the dataset with vibration-induced noise events at the low-energy regime. These characteristic noise events are referred to as \acrfull{lfn} events. Accelerometers were installed to monitor the vibrations, and their measurements were found to be correlated with the detector trigger rate.

\begin{figure}
    \centering
    \includegraphics[width=0.95\linewidth]{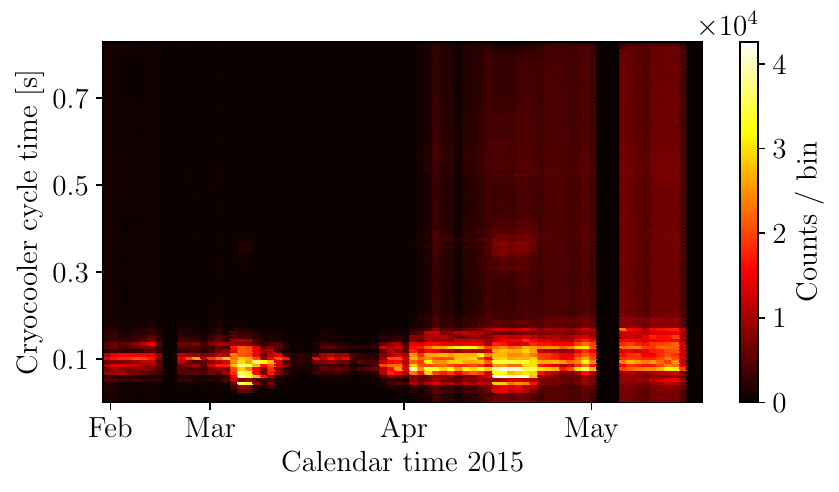}
    \caption{Heatmap of the cryocooler cycle time with a 0.83\,s modulus over the run period in calendar time, with the color intensity representing the number of triggered events. High trigger rates correspond to vibrations occurring 0.1\,s after each cryocooler cycle start. This high trigger rate correlates with low-frequency noise (LFN) in the data, introduced by vibrations from the cryocooler.}
    \label{fig:cryopretime}
\end{figure}

The accelerometer data also indicated that the vibrations were most intense approximately $\sim$0.1\,s after the start of each cryocooler cycle ($\sim$0.83\,s), as shown in Fig.~\ref{fig:cryopretime}, which shows the cryocooler cycle time of each event as a function of the run's calendar time. 

The phonon signal was read out from four individual phonon channels as waveforms (or traces). When the analog sum of the phonon signal exceeded a predefined threshold, the data acquisition (DAQ) system issued a trigger, after which the waveforms were digitized using an \acrfull{adc} and recorded with a digitization rate of 625\,kHz. Each raw trace consisted of 4096 samples, corresponding to a trace length of approximately 6.6\,ms.

The pulse shapes of the majority of the LFN events resemble those of low-energy signal events, posing a significant challenge for their removal from the dataset. Furthermore, the shapes of these noise events vary over time, making them difficult to reject using the traditional pulse-shape cuts, which rely on a reproducible pulse template. The CDMSlite event selection criteria included stringent cuts to remove LFN events, which, due to their similarity to physics pulses, resulted in a reduced signal efficiency. Therefore, this analysis explores a Machine Learning (ML)-based approach to improve the discrimination between the LFN events and real particle interaction events.

The analysis framework developed for CDMSlite Run\,3 is summarized in Sec.~\ref{sec:cdmslite_analysis}. The complete analysis was published in Ref.~\cite{CDMSliter3}. Section~\ref{sec:ML_cut} describes the development of an event selection criterion (also referred to as a cut) based on ML models for the identification of \acrshort{lfn} events, including sample preparation, model building, training, and evaluation. The application of this new cut to the CDMSlite Run\,3 dataset and the calculation of \acrshort{dm} sensitivities to evaluate its performance is presented in Sec.~\ref{sec:cdmslite_reanalysis}. Case studies exploring strategies to navigate the new method's limitations are presented in Sec.~\ref{sec:case_studies}. 

\section{CDMSlite Run 3 analysis}\label{sec:cdmslite_analysis}
The work presented in this paper is based on the analysis framework developed for the CDMSlite Run\,3 presented in Ref.~\cite{CDMSliter3}. This section provides a brief overview of the energy estimation, event selection criteria, and signal efficiencies. In this work, ``signal" refers to pulses created by particle interactions in our detector (i.e., non-LFN events and non-instrumental-noise events), and not necessarily DM candidate interactions unless explicitly stated as ``DM-signal".

\subsection{Energy estimation}
 
\begin{figure}
    \centering
    \includegraphics[width=\linewidth]{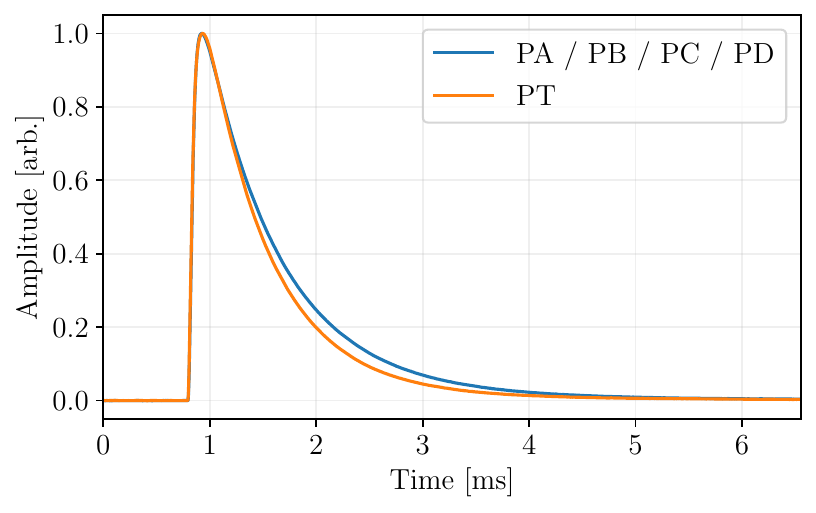}
    \caption{Phonon pulse templates of the individual phonon channels PA, PB, PC, and PD along with the pulse template for the \acrfull{pt} channel (orange). The same template is used for the individual channels (blue).}
    \label{fig:template_channels}
\end{figure}

The \acrfull{of} algorithm \cite{OF1, OF2} was used for reconstructing amplitude, in which the raw traces were fit in the frequency domain to a standard pulse template which represented the expected phonon signal shape to determine the pulse amplitude. The pulse amplitude is representative of the amount of energy deposited in a physics event. The phonon templates for the individual channels and the \acrfull{pt} channel, which is defined as the sum of all four phonon channels, are shown in Fig.~\ref{fig:template_channels}. 
The shape differences between the PT template and the individual channel template arise from the position-dependent pulse shape variations in individual channels being averaged out when summed to form the PT template. The calibration from pulse amplitude to recoil energy was derived from mono-energetic signals caused by the decay of \element{71}{Ge}, which was produced by the activation of \element{70}{Ge} in-situ via neutron capture \cite{activations}.

Corrections to the reconstructed energy were applied to account for variations in the \acrfull{hv} current, the event's position relative to the phonon channels, and changes in the operating temperature.

Two energy definitions were used in the CDMSlite Run\,3 analysis, which were the total phonon energy (in eV$_{\rm t}$) and the electron-equivalent energy (in eV$_{\rm ee}$). The total phonon energy is the sum of the recoil energy and the additional phonon energy produced via the \acrshort{ntl} effect.  The electron-equivalent energy is defined as the electron recoil energy required to produce the observed phonon energy in the detector. This unit was primarily used in the DM search because the backgrounds in the experiment are \acrfull{er}-dominated, i.e., interactions created by particles scattering off of electrons. As event-by-event discrimination between ER backgrounds and the \acrfull{nr} signals of interest was not possible due to the application of HV, all observed energy deposits were calibrated to a known electron-recoil scale.
\subsection{Event selection}\label{sec:cdmsliter3_cuts}
The analysis comprised two categories of cuts: livetime cuts and data quality cuts. The livetime cuts excluded events in non-ideal experimental conditions, such as instabilities in the applied HV, time periods coincident with neutrino beam events from Fermilab \cite{numi}, and issues related to GPS timing. The data quality cuts remove events whose pulse shape is incompatible with the expected signal shape through a series of pulse shape cuts and a radial cut for fiducialization. As this work is part of the data-quality cuts, only this category is discussed in detail in the following sections.

\subsubsection{Pulse shape cuts}
Three main pulse shape cuts are applied to remove instrumental noise events. These include 1) electronic glitch events, which are characterized by sharp rise and fall times, 2) square-pulse events, which are characterized by a fast rise time followed by a plateau region and a fast fall time, and 3) vibration-induced \acrshort{lfn} events, which generally exhibit slower rise times compared to signal events. The pulse shape discrimination is based on two templates: a standard pulse template which was developed for energy reconstruction (see Fig. \ref{fig:template_channels}) and a template which represents the shape of the particular noise events. The difference in the OF \chisq{} values obtained from these two templates was calculated, and a cut was placed in the energy–\deltachisq{} plane to remove these events \cite{CDMSliter3}. This selection is commonly referred to as the \deltachisq{}{} cut. 

\subsubsection{Low-Frequency Noise cuts}
The selection criterion used for removing the \acrshort{lfn} events consisted of two separate cuts: 1) for event energies above 5\,\kevt{}, the \deltachisq{} cut was applied. 2) for events below 5\,\kevt{}, it was difficult to remove these events with only a pulse shape cut, thus a bifurcated analysis incorporating additional parameters, such as trigger correlation and crycooler cycle time (shown in Fig.~\ref{fig:cryopretime}), was used to remove LFN events. A detailed description of this bifurcated analysis is provided in Ref.~\cite{CDMSliter3}. 

In brief, the bifurcated cut consists of two uncorrelated sets of selection criteria. One of the sets uses the \deltachisq{} parameter with three different templates that represent typical \acrshort{lfn} pulse shapes. The other set uses cryocooler cycle time and correlations between detectors arising from vibrational noise observed simultaneously across multiple detectors. This method was effective in removing the LFN events from the dataset, however, this approach reduced the signal efficiency by about an order of magnitude for energy deposits close to the experimental threshold. The motivation of this work is therefore to develop a new cut that effectively removes LFN events while preserving a high signal efficiency.

\subsubsection{Radial cut}\label{sec:radial_cut}
Non-uniformities in the applied electric field induce non-uniform NTL amplification within the detector. Events occurring at larger radii experience smaller NTL gain due to the reduced potential difference compared to events in the inner regions or the bulk, and are therefore reconstructed at systematically lower energies. Thus, these so-called reduced-NTL (RNTL) events are removed from the energy spectrum to ensure a uniform detector response model for signal-like events.

Since the true interaction position cannot be determined, a radial parameter is constructed from the fast and slow components of the phonon pulse template and the pulse start-time difference between the inner (B, C, and D) channels and outer phonon channel (A). The fast component, which is associated with the pulse rise time, carries information regarding the event position, and the slow component, which is associated with the pulse fall time, corresponds to the measured energy. For removing the RNTL events, a cut is applied in the radial parameter vs. energy plane. This radial cut imposes an analysis threshold of 70\,\evee{}. Further details can be found in Ref.~\cite{CDMSliter3}. 

\subsection{Efficiencies}\label{sec:cdmsliter3_efficiencies}
The signal efficiency estimates the fraction of signal events surviving the event selection criteria. The final DM-signal efficiency as a function of energy consists of three main factors: (1) the trigger efficiency, which quantifies the fraction of events that make it past the hardware trigger, (2) data quality cuts, which determine the fraction of events that survive cuts such as glitch, square-pulse, and LFN cuts, and (3) the radial cut, which removes events at larger radii.

The hardware trigger efficiency was evaluated using calibration events that triggered in other active detectors. For events with a measurable energy deposition in the CDMSlite detector, the trigger efficiency was calculated as a function of energy from the fraction of events that also triggered the CDMSlite detector. The data quality cut efficiency was determined by generating raw traces of signal-like events and passing them through the data quality cuts and finding the fraction of events that pass these cuts as a function of energy.

The radial parameter efficiency was derived using Monte-Carlo simulations \cite{cdmsliter2,CDMSliter3}. 
The radial parameter vs. energy cut was then applied to these simulated events, yielding the radial efficiency. This efficiency is subsequently multiplied by the fraction of non-RNTL events ($\sim$86\%) which is referred to as NTL efficiency to obtain the final radial cut efficiency. Further details can be found in Ref.~\cite{CDMSliter3}.

To obtain the DM-signal efficiency, the hardware trigger efficiency was first convolved with the data quality cut efficiency, and the result was then convolved with the radial cut efficiency.  The efficiencies were then parameterized using a functional form as shown in Eq. \ref{eq:final_signal_efficiency}.
\begin{equation}
f(E) = \mu_1 \cdot \left(1-{\rm erf}\left(\frac{E-\mu_2}{\sqrt{2}\cdot\mu_3}\right)\right),
\label{eq:final_signal_efficiency}
\end{equation}
where $E$ is the energy, and $\mu_1$, $\mu_2$, and $\mu_3$ are the parameters to be fitted. The efficiencies for the CDMSlite Run\,3 analysis and the modified analysis presented in this work are described in Sec.~\ref{sec:cdmsliter3_efficiencies_new}. 

\subsection{Energy resolution model}
The resolution model developed for this analysis consists of three main components: (1) the baseline resolution arising from the electronics and experimental conditions, (2) position dependence, which distorts the pulse shape and introduces uncertainties in the energy reconstruction, and (3) the Fano factor, which handles the uncertainty in the charge production. The detector response model was parameterized as a function of the recoil energy (in electron-equivalent units) E$_{\rm ee}$ such that total energy resolution $\sigma_{\rm T}$(E$_{\rm ee})$ is written as follows:

\begin{equation}
    \sigma_{\rm T}(E_{\rm ee}) = \sqrt{\sigma^2_{\rm E} + \sigma^2_{\rm F}(E_{\rm ee}) + \sigma^2_{\rm PD}(E_{\rm ee}}),\\
        \label{eq:energy_res1}
\end{equation} 
where $\sigma^2_{\rm E}$ is the energy independent baseline resolution. $\sigma_{\rm F}(E_\text{ee})$ is the Fano resolution term, which is proportional to $\sqrt{E_{\rm ee}}$ and $\sigma_{\rm PD}(E_\text{ee})$ is the position-dependence term, which is proportional to $E_{\rm ee}$. The fit parameters and further details can be found in Ref.~\cite{CDMSliter3}.

\section{ML-based LFN identification}\label{sec:ML_cut}
The identification of \acrshort{lfn} events in the dataset is particularly critical at low energies, where their amplitudes overlap with those of signal events. Consequently, the classification is performed as a function of amplitude (proportional to energy) with particular emphasis on improving the performance in the low-amplitude region. The goal is to identify and remove LFN events without compromising the signal efficiency, using ML-based classifiers. 

\subsection{Sample preparation}
The input to the classifier consists of raw traces, labeled as 1 for signal and 0 for LFN-background. There are three phonon channel configurations considered for the classification task: (1) single-channel analysis, in which each phonon channel is evaluated independently, (2) the use of a \acrfull{pt} channel, and (3) a multi-channel approach, in which all four individual phonon channels are considered simultaneously.

\subsubsection{Signal sample preparation}

The principal idea of signal sample generation is to add template-based or simulated phonon pulses onto baseline noise traces. These baseline traces are extracted from a randomly triggered dataset, which are then passed through pulse rejection algorithms to remove traces with pulses. 

\begin{center}
    \underline{Single- and PT-channels}\\
\end{center}

Phonon pulse templates for individual channels and for the PT-channel were developed as part of the CDMSlite analysis, which are shown in Fig.~\ref{fig:template_channels}. Signal samples for both single-channel and PT-channel configurations are generated by adding these templates scaled to the desired amplitude to baseline noise traces from data.

The pulse shapes observed in the individual channels vary based on the event position within the detector. For instance, a channel closer to the interaction typically exhibits not only a larger amplitude from the high number of phonons reaching the channel, but more importantly a faster rise time compared to other channels and vice versa. The radial cut developed for the CDMSlite Run\,3 analysis exploits these position-dependent pulse shape differences to reject events originating at larger radii, as described in Sec.~\ref{sec:cdmsliter3_cuts}. As a result of these position-dependent pulse shape differences, a single template is not fully representative of the pulse shapes in any given channel.
In contrast, the PT-channel trace, which is the sum of all four phonon channels traces, averages over the position-dependent variations in pulse shape. This averaging improves the robustness of the energy reconstruction and potentially helps the identification of \acrshort{lfn} events compared to the single-channel configuration.
\newpage
\begin{center}
    \underline{Multi-channel}\\
\end{center}

The multi-channel configuration utilizes information from all four phonon channels simultaneously for the event classification. Since the phonon pulse template is identical for all four channels, it cannot be directly used to generate realistic signal-like events due to the lack of the pulse shape variations arising from the position dependence within the detector.

A more realistic modeling of the pulse shapes with a simulation-based approach is required. As the original simulation framework for this experiment is no longer available, a novel method is developed to construct pulses corresponding to different event positions within the detector.

As discussed previously, pulse shapes vary with the interaction position in the detector volume. Therefore, the simulation must incorporate this position dependence using events that are uniformly distributed throughout the detector. \element{71}{Ge} activation peaks provide a suitable dataset for this, as they are expected to be homogeneously distributed within the detector volume. Among the activation peaks, the K-shell peak has the highest statistics and is therefore used for the simulation. The CDMSlite Run\,3 measured a K-shell peak energy of 10.354$\pm$0.002\,\kevee{} with a peak resolution of 108$\pm$2.0\,\evee{} \cite{CDMSliter3}. The events which pass all the quality cuts that fall within $\pm$5$\sigma$ of the K-shell peak energy are selected.

\begin{figure}
    \centering
    \includegraphics[width=0.9\linewidth]{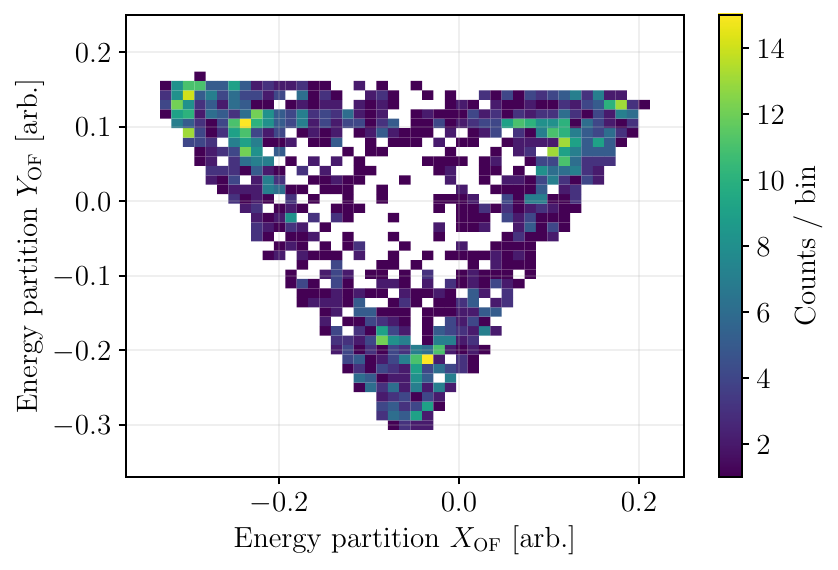}
    \caption{Partition plot of K-shell events, which is a representation of the localization of the events in the $x$-$y$ plane, using the energy partition information calculated using Eqs.\,\ref{eq:Xpartition} and \ref{eq:Ypartition}.}
    \label{fig:K_shell_partition}
\end{figure}

Although the true interaction position cannot be directly determined, the position can be localized using the relative energy deposition across the wedge-shaped inner phonon channels (see Fig.~\ref{fig:detector}). This positional localization is referred to as the ``energy partition", which is defined as follows:

\begin{equation}
   X_\text{OF} = \frac{\cos(30^{\circ}){D}_\text{OF} + \cos(150^{\circ}){B}_\text{OF} + \cos(270^{\circ}){C}_\text{OF}}{{D}_\text{OF} + {B}_\text{OF} + {C}_\text{OF}},
    \label{eq:Xpartition}
\end{equation}
\begin{equation}
   Y_\text{OF} = \frac{\sin(30^{\circ}){D}_\text{OF} + \sin(150^{\circ}){B}_\text{OF} + \sin(270^{\circ}){C}_\text{OF}}{{D}_\text{OF} + {B}_\text{OF} + {C}_\text{OF}},
    \label{eq:Ypartition}
\end{equation}
where $B_\textnormal{OF}$, $C_\textnormal{OF}$, and $D_\textnormal{OF}$ are the OF amplitudes of channels PB, PC, and PD, respectively. Plotting the $X_\textnormal{OF}$ and $Y_\textnormal{OF}$ energy partitions, called a partition plot, serves as a tool to assess the spatial distribution of events in the $x$-$y$ plane. The energy partitions of the selected K-shell events are shown in Fig.~\ref{fig:K_shell_partition}, where the symmetric population among the corners corresponding to the three wedge-shaped phonon channels is consistent with the expected homogeneity of bulk events within the detector. 

This novel simulation method fits the K-shell peak pulses from the four phonon channels with an exponential function with two rise time components and two fall time components. The fit function $y$ as a function of time $t$ is given by
\begin{align}
y (t) = a(e^{-(t+t_0)/\tau_{f_1}} + e^{-(t+t_0)/\tau_{f_2}} \nonumber\\- e^{-(t+t_0)/\tau_{r_1}} - e^{-(t+t_0)/\tau_{r_2}}),
    \label{eq:exp_fit_2r2f}
\end{align}
where $a$ is the amplitude, $\tau_{f_1}$ and $\tau_{f_2}$ are the fall time constants, $\tau_{r1}$ and $\tau_{r2}$ are the rise time constants, and $t_0$ is the offset in time with respect to the trigger point.

Each event has traces from four phonon channels, resulting in a total of 16 fit parameters. These parameters are expected to be correlated with the event position in the detector, given by the partition variables $X_\text{OF}$ and $Y_\text{OF}$. A decision tree regressor is trained using the fit parameters as inputs and the corresponding partition variables as targets. The training dataset consists of 1400 samples. The trained regressor is then used to predict the fit parameters for arbitrary values of $X_\text{OF}$ and $Y_\text{OF}$.

\begin{figure*}
    \centering
    \includegraphics[width=\linewidth]{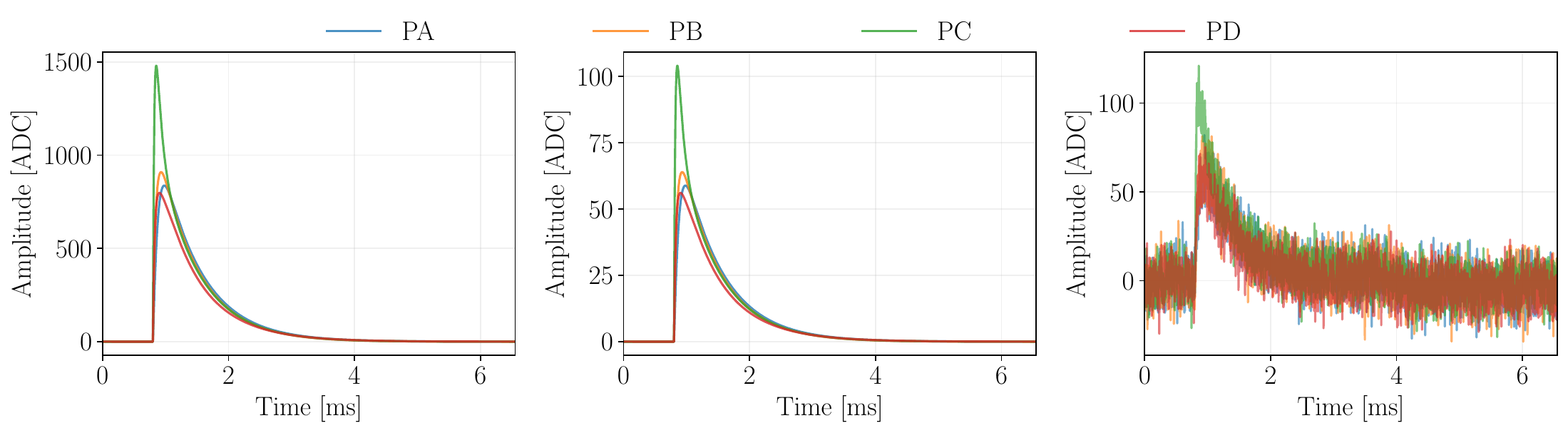}
    \caption{Traces generated from the predicted fit parameters using the SUPR method for the four phonon channels PA-PD are shown on the left, and the traces rescaled to the desired PT amplitude in the middle. On the right, baseline noise is added to the rescaled traces.}
    \label{fig:SUPR_pred_scale_rescale}
\end{figure*}

The input partition $X_\text{OF}$ and $Y_\text{OF}$ values are sampled from a uniform distribution over the triangular region in the partition plot, which are passed through the trained regressor. The predicted fit parameters are used to reconstruct pulse shapes, which are then scaled based on the desired PT amplitude. The rescaled pulses are then added to baseline noise traces to generate simulated signal events. Figure~\ref{fig:SUPR_pred_scale_rescale} illustrates example pulses of an event generated using this approach, which are rescaled and added to baseline noise to produce simulated signal events used in training the multi-channel LFN classifier.

This method where the pulse shapes are modeled based on position using a regression framework is referred to as \acrfull{supr}. The SUPR method serves as a proof of principle and can be further fine-tuned to improve performance.

\subsubsection{LFN-background preparation}
As the \acrshort{lfn} events cannot be simulated using a simulation framework due to their varying shape, the sample preparation for the LFN-background is data-driven. The CDMSlite Run\,3 events are selected during the periods when the vibrations are intense, specifically between March 4th and 5th, and between April 2nd and 30th. These events are passed through the full set of analysis cuts developed for the CDMSlite Run\,3 analysis except the LFN cut. Events that pass all the cuts are then passed through the previously developed LFN cut, and those that fail the LFN cut are used as the LFN-background sample for training and testing the classifier. It should be noted that there is a non-zero possibility of signal contamination in this sample. However, the selected events are predominantly LFN, thus the resulting dataset is an LFN-enriched sample. 

\subsection{Model building}

The \acrshort{ml} models used for the identification of \acrshort{lfn} events are \acrfullpl{cnn}  \cite{CNN1}  and \acrshortpl{cnn} with \acrfull{lstm} \cite{LSTM} architectures. The CNN models are chosen to utilize differences in the pulse shape between signal and LFN-background events, while the addition of LSTMs is motivated by its effectiveness in capturing temporal dependencies in time-series data.

Each architecture is applied to the three different channel configurations: single-channel, PT-channel, and multi-channel. The training sample size varies for these channel modes, in particular, the single-channel configuration provides four times as many training samples (one per channel, i.e., four samples per event) compared to the PT and multi-channel configurations (one per event). 

Hyperparameter optimization, which tunes the ML model's parameters, was performed separately for the three different channel configurations so the architectures and the parameters slightly vary between the three configurations. Appendix~\ref{appendix:cnn} details the model architectures and their optimized parameters. 

Several standard evaluation metrics were used during the classifier development such as \acrfull{roc} curves and confusion matrices; however, the two primary metrics used for evaluating performance for this classification are signal efficiency (recall) and signal purity (precision), which are defined as function of the amplitude $a$:
\begin{equation}
    \text{Efficiency(\textit{a})} = \frac{\text{TP(\textit{a})}}{\text{TP(\textit{a})}+\text{FN(\textit{a})}},
    \label{eq:sig_efficiency}
\end{equation}
\begin{equation}
    \text{Purity(\textit{a})} = \frac{\text{TP(\textit{a})}}{\text{TP(\textit{a})}+\text{FP(\textit{a})}},
    \label{eq:sig_purity}
\end{equation} 
where TP (True Positives) denote the number of signal events correctly identified. FN (False Negatives) and FP (False Positives) denote signal and LFN-background events that are incorrectly classified, respectively.

The efficiency is evaluated using only the signal sample, i.e. TP and FN which correspond to the number of signal events that are correctly and incorrectly identified, respectively. In contrast, the purity is calculated using both correctly identified signal events (TP) and misclassified LFN-background events (FP), and therefore depends on a mixture of signal and LFN-background samples. The amplitude distributions of signal and background samples differ, which can introduce features in the purity when calculated as a function of amplitude. The signal sample amplitudes are uniformly distributed. However, the LFN-background amplitude distribution peaks around 12\,ADC units for individual channels and 22\,ADC units for the PT-channel. This leads to a feature resembling a ``dip"  observed in the purity calculation as will be seen in Fig.~\ref{fig:cnn_all_channel} and Fig.~\ref{fig:cnn_cnnlstm_all_channel}. This dip does not impact the interpretation of the signal purity and instead reflects an intrinsic feature of the underlying data.

In the classification task, the purity is given more importance than signal efficiency, as purity directly reflects the level of LFN-background rejection. The LFN-background rejection is essential as increased background contamination in the energy spectra leads to weaker sensitivities, particularly for limit-setting methods that treat all remaining events after the cuts as DM-signals, such as the \acrfull{oi} method \cite{OI_Yellin}. Even for methods such as \acrfull{plr}, where background models are incorporated, the background model could not be constructed for LFN-background events, requiring stringent cuts to remove them.  Thus, LFN-background rejection is treated as a baseline requirement, with improved signal efficiency as the final goal. 

The training sample size for the single-channel configuration consists of 80,000 samples, with equal proportions of signal and LFN-background events. This sample size is primarily constrained by the available computational resources for model training. For the PT- and multi-channel configurations, the training sample size is 40,000 samples each, also with equal proportions of signal and LFN-background events.

\subsubsection{Convolutional Neural Network}
% \raggedbottom
The CNN model consists of two 1D convolutional layers with Rectified Linear Unit (ReLU) activation \cite{relu} using a large kernel size to effectively capture local dependencies. The option ``padding" is used to regulate the output length of the feature map, and is set to ``same'' to maintain the input dimensions after convolution. A \acrfull{gap} \cite{GAP,GAP_in_cnn} layer is used to reduce dimensionality instead of flattening, as the resulting weighted features are later used for calculating \acrfullpl{cam} \cite{CAM}. This is followed by fully connected (dense) layers, with the final dense layer using a ``sigmoid" activation function  \cite{goodfellow2016DNN} with a single neuron  to produce binary probabilistic predictions. Outputs closer to 0 correspond to LFN-background-like events, and closer to 1 correspond to signal-like events.

\begin{figure*}
    \centering
    \includegraphics[width=\linewidth]{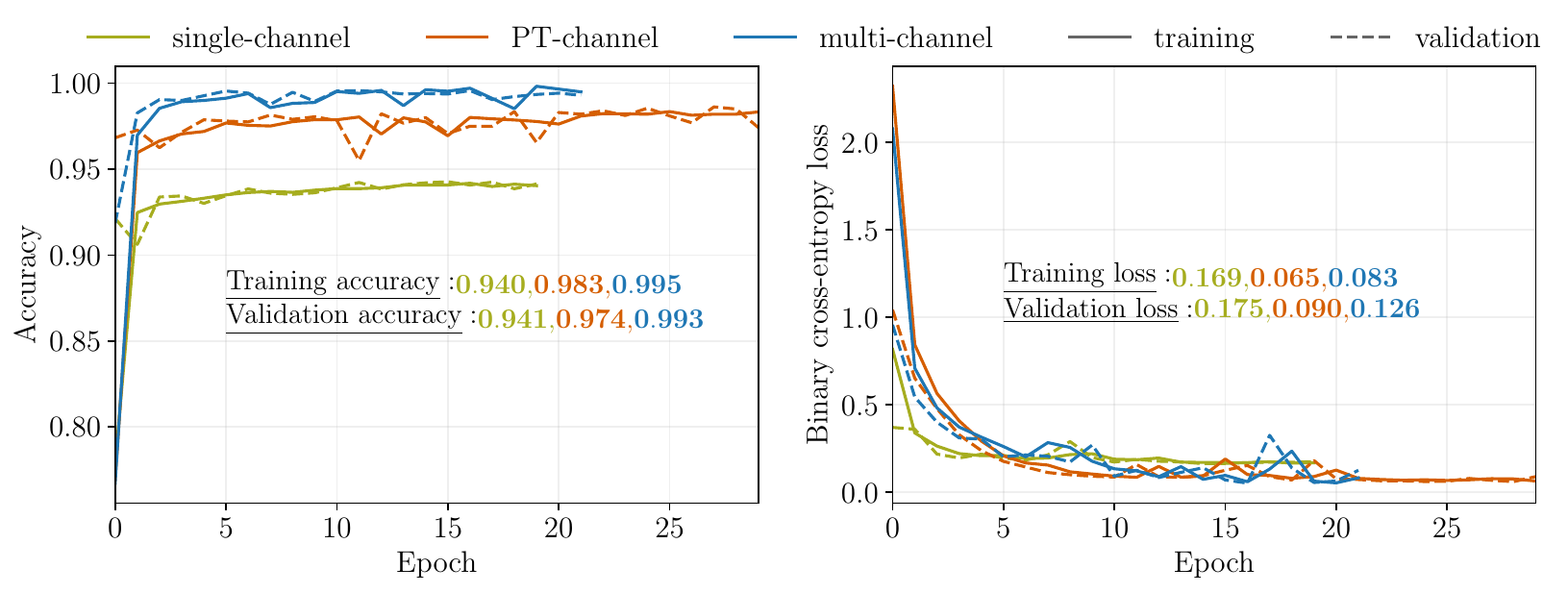}
    \caption{Training and validation accuracy (left) and loss curves (right) of the classifier training plotted against epochs for single-(green), PT-(red), multi-(blue) channel configurations, where the number of epochs is determined by early stopping. The training set is shown in solid and the validation is shown in dashed lines, with the accuracy and loss of each configuration provided in the text.}
    \label{fig:all_channel_training}
\end{figure*}

The model is trained using the Adam optimizer \cite{adam} with a learning rate of 0.001 and a binary cross-entropy loss function \cite{bce_loss}. Early stopping is applied during training to mitigate overfitting.

Figure~\ref{fig:all_channel_training} presents the training of the CNN classifier for single-, PT-, and multi-channel configurations, where the accuracy and loss curves are shown as a function of epochs. Early stopping was used during the training of the classifiers to mitigate over-training. The single-channel model was trained for 19\,epochs, the PT-channel model was trained for 29\,epochs, and the multi-channel model was trained for 21\,epochs. The multi-channel model achieves a higher training and validation accuracy compared to the single- and the PT-channel. The PT-channel model has the lowest loss compared to the other two models.

The single-channel test dataset comprises 131,983 samples of which 37,983 are signal events and 94,000 LFN-background events. These samples were not used during training and were used to evaluate the performance of the network. The PT-channel test dataset consists of 27,178 samples, of which 3,678 are signal events and 23,500 are LFN-background events. The multi-channel test dataset contains 26,086 samples, comprising 6,487 signal events and 19,599 LFN-background events.

\begin{figure}
    \centering
    \includegraphics[width=1\linewidth]{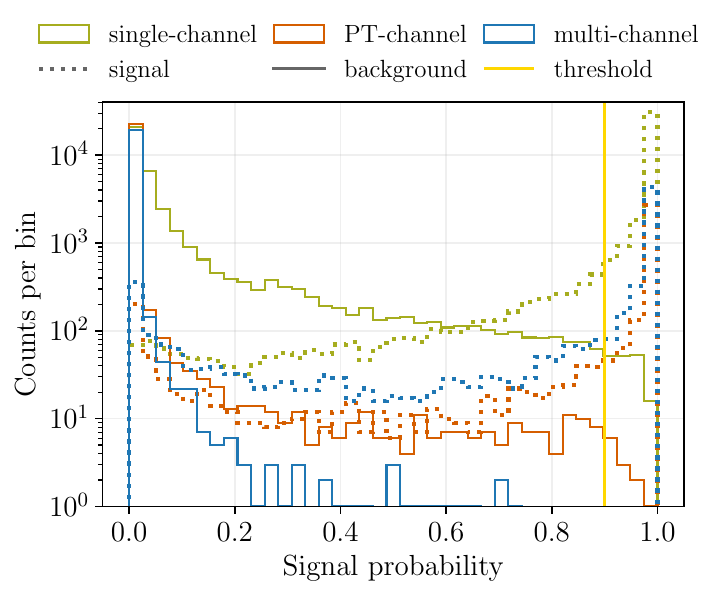}
    \caption{Signal probability distribution for the CNN classifier for single-(green), PT-(red), multi-(blue) channel configurations tested on unseen test samples. The LFN-background sample and the signal sample are shown in solid lines and dotted lines, respectively. The signal probability threshold of 0.9 is shown by the yellow solid line. }
    \label{fig:all_channel_score}
\end{figure}

The network output, which ranges between 0 and 1, is referred to as the ``signal probability" in this analysis. The distributions of the signal probability for the single-, PT-, and multi-channel test samples are shown in Fig.~\ref{fig:all_channel_score}. Signal and LFN-background events are peaked at 1 and 0, respectively, and events with intermediate outputs are of particular interest. During the classifier development, these intermediate cases identified in early model iterations were examined in detail to optimize the model architecture and inform the sample preparation, with the aim of minimizing misclassification.

A threshold on the signal probability is used to make the final binary classification as either signal or LFN-background. Applying a threshold on the signal probability involves a trade-off: increasing the threshold reduces the probability of LFN-background events being misidentified as signal, thereby improving purity, but it also reduces signal efficiency. An optimal threshold is reached when a further increase in the threshold no longer yields a net benefit, as the purity gain no longer compensates for the signal efficiency loss. A threshold of 0.9 was chosen, as multiple network architectures evaluated during the development converged to this value balancing signal efficiency and purity.

\begin{figure}
    \centering
    \includegraphics[width=0.9\linewidth]{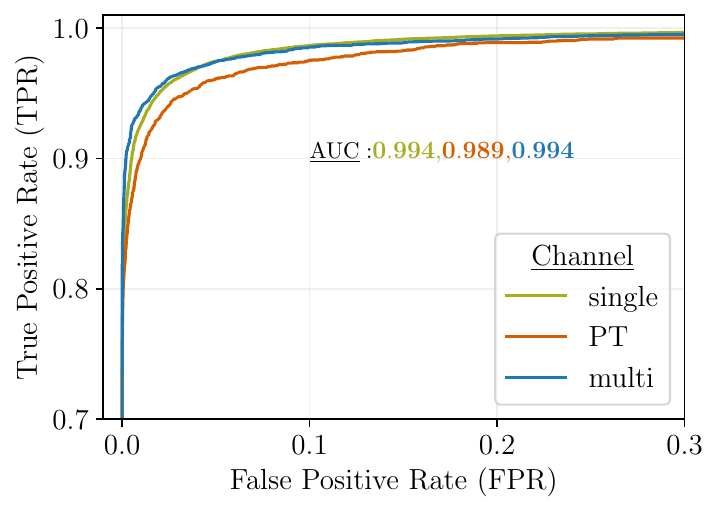}
    \caption{\acrshort{roc} curves for the CNN classifier for single-(green), PT-(red), multi-(blue) channel configurations tested on unseen test samples. The AUC of the ROC is shown in the text.}
    \label{fig:all_channel_roc}
\end{figure}

Figure~\ref{fig:all_channel_roc} shows \acrshort{roc}s derived from the test samples for the three channel configurations.
The \acrshort{roc} curve illustrates performance of a binary classifier, with an ideal classifier corresponding to the top-left corner of the plot, where the true positive rate (TPR) is 1 and false positive rate (FPR)  is 0. The \acrfull{auc}, which indicates the model’s ability to distinguish between the two classes with higher values indicating better separability, is 0.993 for the single-channel model, 0.992 for the PT-channel model, and 0.981 for the multi-channel model. 

\begin{figure}
    \centering
    \includegraphics[width=\linewidth]{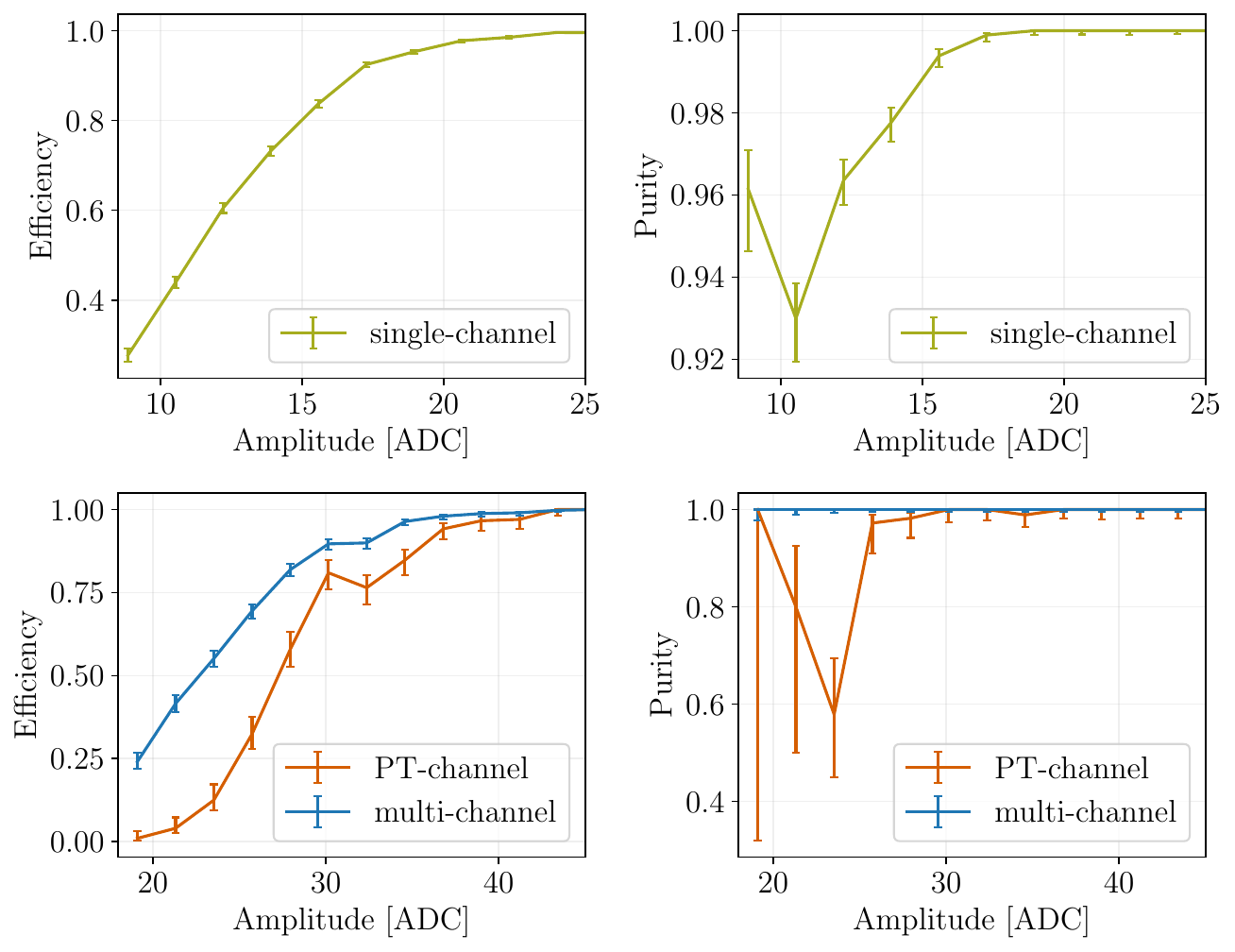}
    \caption{Signal efficiency (left) and purity (right) of the single-channel classifier (top) and PT- and multi-channel (bottom) calculated with the test samples. The signal probability threshold is placed is placed at 0.9. The error bars denote the statistical uncertainty.}
    \label{fig:cnn_all_channel}
\end{figure}

The signal efficiency and purity for the single-channel classifier and for the PT- and multi-channel classifiers for a signal probability threshold at 0.9 are shown in Fig.~\ref{fig:cnn_all_channel}.  The signal efficiency for the single-channel configuration is shown separately from the PT- and multi-channel configurations due to differences in the x-axis. The single-channel results are based on the amplitude of individual channel traces, whereas the PT- and multi-channel results use the amplitude of the PT trace. Both efficiency and purity decrease with decreasing amplitude. It can be observed that the multi-channel configuration achieves overall better signal efficiency and purity compared to the single- and PT-channel configurations. Importantly, the multi-channel model reaches 100\% purity on the test sample with a statistical uncertainty of up to 2.25\%, making it the final choice for this analysis. The dip observed in the purity distribution arises from the non-uniform amplitude distribution of the LFN-background sample, as discussed previously.

\acrfullpl{cam} \cite{CAM} are used to identify which features extracted by the convolutional layers contribute most significantly to the classification decision. To allow this, a GAP layer was introduced between the final convolutional layer and the fully connected part of the network.

For an output of the convolution layer $A_i$ of the $i^\text{th}$ filter, the \acrshort{cam} as a function of time $t$ is calculated as:
\begin{equation}
    \text{CAM}(t) = \sum_{i}w_i\cdot A_i(t),
    \label{eq:CAM}
\end{equation}
where $w_i$ is the weight of the $i^\text{th}$ filter. 

\begin{figure}
     \centering
     {
         \includegraphics[width=0.45\linewidth]{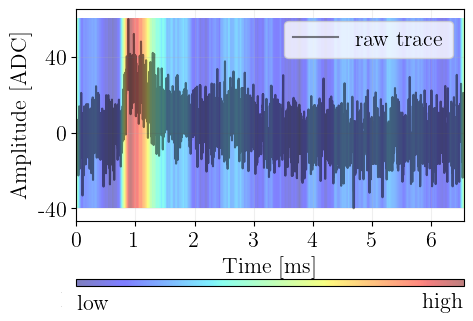}
         }
     % \
     {
         \includegraphics[width=0.45\linewidth]{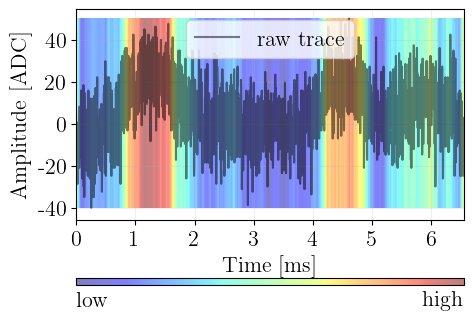}
        }
        \\
     % \
     {
         \includegraphics[width=0.45\linewidth]{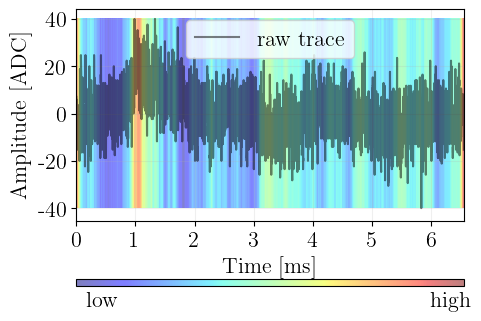}
         }
     % \
     {
         \includegraphics[width=0.45\linewidth]{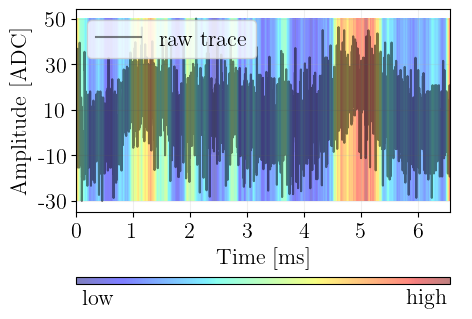}
         }
     \caption{Activation maps of randomly selected traces of each class (top left: TP, top right: TN, bottom left: FP, bottom right: FN) of the single-channel classifier with a signal probability threshold of 0.9. The raw traces are shown in transparent black over which the color maps representing the feature importance for the prediction are overlaid. The activation maps are calculated using Eq.~\ref{eq:CAM}.}
     \label{fig:CAM}
\end{figure}

When overlaid onto the input trace, these activation maps highlight the regions that contribute most to the model's decision. During the initial stages of classifier development, CAMs were systematically examined for the misclassified events and iteratively used to improve the model architecture and sampling of baseline noise data to add pulses for the signal sample generation. The CAMs corresponding to \acrshort{tp}, \acrshort{fp}, \acrshort{fn}, and \acrfull{tn} are shown in Fig.~\ref{fig:CAM}.

For the TP event, the CAM highlights larger weights associated with the sharp rise time of the pulse as opposed to the TN event which exhibits a secondary larger set of weights at $\sim$5\,ms along with a more distributed weight structure across the entire trace. Misclassified signal (FN) events show broader weight distributions along the trace, especially also at $\sim$5\,ms which contributes to their incorrect classification. Misclassified LFN-background (FP) events exhibit larger weights around the trigger window (set at the 500th time bin, i.e. 0.8\,ms), where the sharp rise-time of pulses is expected.  In this region, noise features can mimic signal-like characteristics, leading to misidentification.

Overall, CAMs are a useful tool for probing the CNN models which not only guide improvements in the dataset construction and model performance but also provide insight into the features used by the network for classification.

For all three channel configurations, pre-processing of the raw traces, such as low-pass filtering and moving averaging, were explored during CNN training and evaluation. Among the three configurations, the PT channel showed better performance with low-pass filtering with improvement in both accuracy and purity than using unprocessed traces. Consequently, low-pass filtering is applied only to the PT-configuration.

\subsubsection{Convolutional Neural Network with Long Short-Term
Memory}

The CNN-LSTM model combines a 1D CNN with LSTM layers, and consists of two convolutional layers with large kernel size to effectively capture local dependencies. Following the convolutional layers, there are two LSTM layers inter-layered with Batch normalization \cite{batchnormalization} for the PT and multi-channel models. Batch normalization was omitted from the single-channel model, as hyperparameter optimization identified a configuration without it that yielded higher accuracy. A dense layer with the same units follows the second LSTM layer. The final output layer consists of a single unit with a sigmoid activation function to produce a signal probability. After each layer, Dropout \cite{dropout} layer, which drops the weights randomly, is used to mitigate overfitting. The model is compiled using the Adam optimizer with binary cross-entropy loss, similar to the CNN architecture. The network architecture varies between the single-, PT-, and multi-channel configurations, as hyperparameter optimization was performed independently to obtain the best-performing model for each case. Details of the parameters are provided in Appendix~\ref{appendix:cnnlstm}.

\begin{figure}
    \centering
    \includegraphics[width=\linewidth]{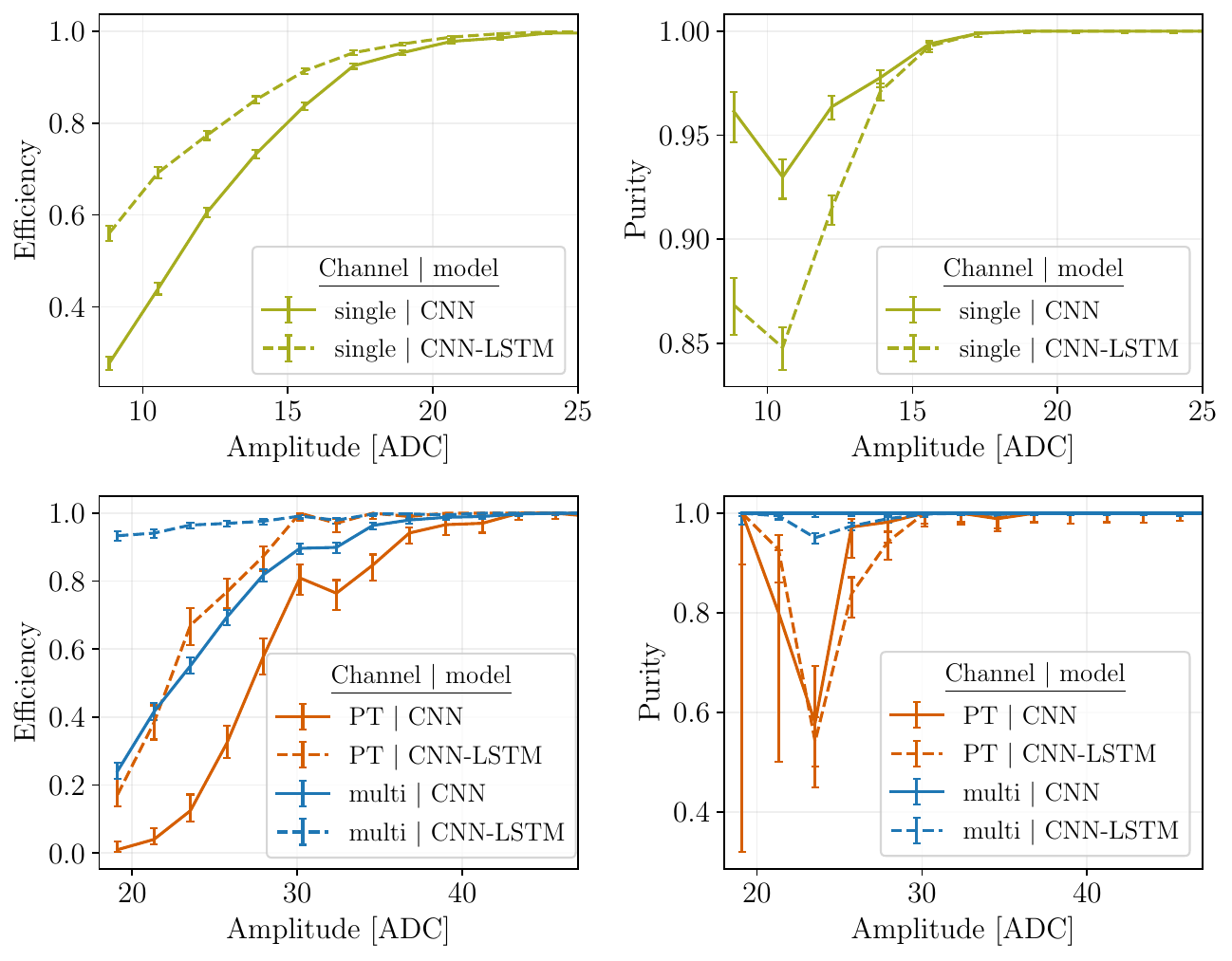}
    \caption{Performance of the CNN (solid) and the CNN-LSTM (dashed) for the LFN classification. The efficiency (left) and purity (right) as functions of the amplitude are calculated for a signal probability threshold at 0.9, with error bars showing the statistical uncertainty.}
    \label{fig:cnn_cnnlstm_all_channel}
\end{figure}

The efficiency and purity of the single-, PT- and multi-channel configuration using the CNN-LSTM is shown in Fig.~\ref{fig:cnn_cnnlstm_all_channel} along with the CNN model results. Overall, the CNN-LSTM classifiers achieve higher signal efficiency compared to the CNN models across all three channel configurations. However, the CNN models outperform the CNN-LSTM classifiers in terms of signal purity, which is prioritized in this analysis.

The improved signal efficiency of the CNN-LSTM models can be attributed to the LSTM's ability to utilize temporal correlations in the data. Signal events exhibit a well-defined temporal structure characterized by the sharp rise time followed by a fall lasting for $\sim$1.5\,ms, In contrast, LFN-background traces exhibit more variations in the temporal structure, making them less consistently identifiable. This variability poses a challenge in LFN-background rejection using LSTMs and consequently lowers the signal purity compared to the CNN models.

In conclusion, a multi-channel configuration using a CNN classifier was selected for the identification of LFN-background due to its 100\% purity on the test dataset.

\section{CDMSlite Run 3 reanalysis}\label{sec:cdmslite_reanalysis}

\begin{figure}
    \centering
    \includegraphics[width=\linewidth]{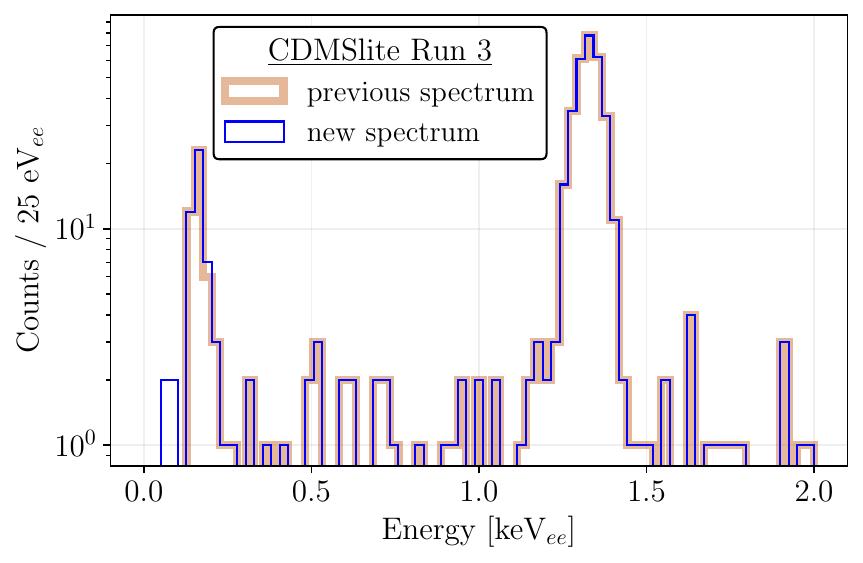}
    \caption{Final energy spectra of CDMSlite Run\,3 using the previous analysis with the bifurcated LFN cut (orange), and the new analysis with the CNN-based LFN cut (blue).} 
    \label{fig:final_spectra_noPull}
\end{figure}

The multi-channel classifier with a signal probability threshold of 0.9 is applied to the CDMSlite Run\,3 dataset, which consists of $\sim$24.7M physics-triggered events. The fraction of events passing the previous LFN cut and the new CNN-based LFN cut is 0.26\% and 0.34\%, respectively. All events that passed the previous LFN cut also passed the newly developed CNN-based LFN cut. After applying all other quality and livetime cuts, including the radial cut, the number of events remaining within the \acrfull{roi} between 0.07\,\kevee{} and 2\,\kevee{} is 401 for the previous analysis and 406 for the new analysis. The corresponding energy spectra are shown in Fig.~\ref{fig:final_spectra_noPull}, where the newly identified events are seen as the differences between the two analyses visible in specific bins at energies between 70\,\evee{} to 190\,\evee{}.

\subsection{Signal efficiencies}\label{sec:cdmsliter3_efficiencies_new}

It is important to note that the signal samples used to calculate the signal efficiencies in the previous analysis are not identical to those used in the current analysis using the CNN-based LFN cut. As the original signal samples used in the previous analysis are no longer available, a similar method for generating signal sample traces was adopted. The principal idea is to construct signal traces from baseline noise traces by adding signal templates whose pulse amplitudes are scaled to desired energies. The signal pulse templates are derived from the \element{71}{Ge} K-shell activation peak as described earlier. As different samples were used to calculate the efficiencies for the previous and new method, although both samples were prepared using the same methodology, the results presented in this section should be interpreted with caution. Despite this, the results serve as a proof of principle for the CNN-based LFN cut and motivates future work to incorporate this approach.

\begin{figure}
    \centering
    \includegraphics[width=\linewidth]{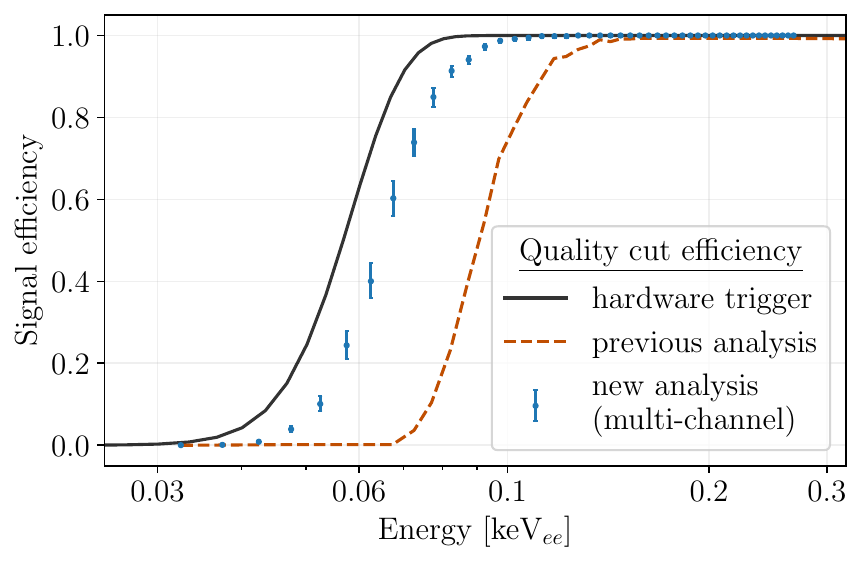}
    \caption{Signal efficiency as a function of energy for the quality cuts that include the CNN-based LFN cut in blue and the previous analysis which includes the bifurcated LFN cut shown in orange (dashed) with the hardware trigger efficiency shown in black, taken from Ref.~\cite{CDMSliter3}.}
    \label{fig:DQ_efficiency_old_new}
\end{figure}
\begin{figure}
    \centering
    \includegraphics[width=\linewidth]{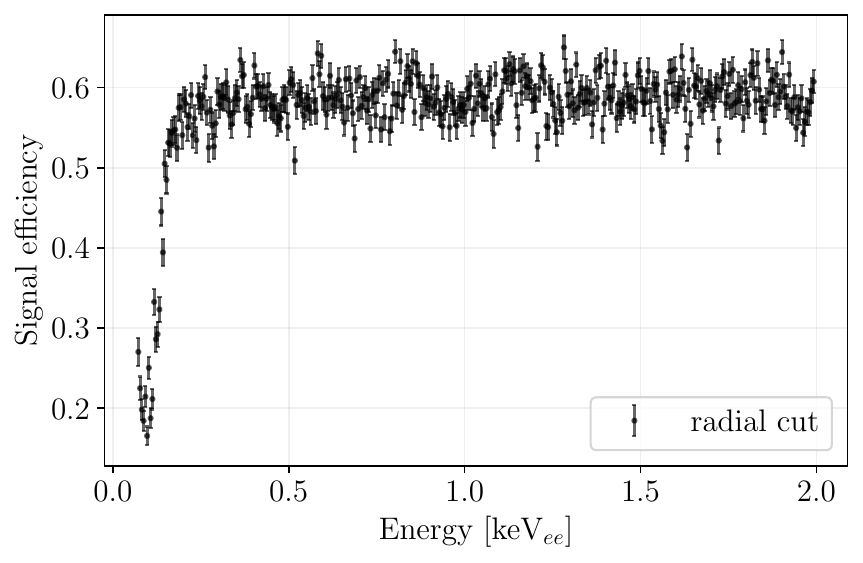}
    \caption{Radial cut efficiency developed for the CDMSlite Run\,3 analysis.~\cite{ryanthesis} }
    \label{fig:radial_efficiency}
\end{figure}
\begin{figure}
    \centering
    \includegraphics[width=\linewidth]{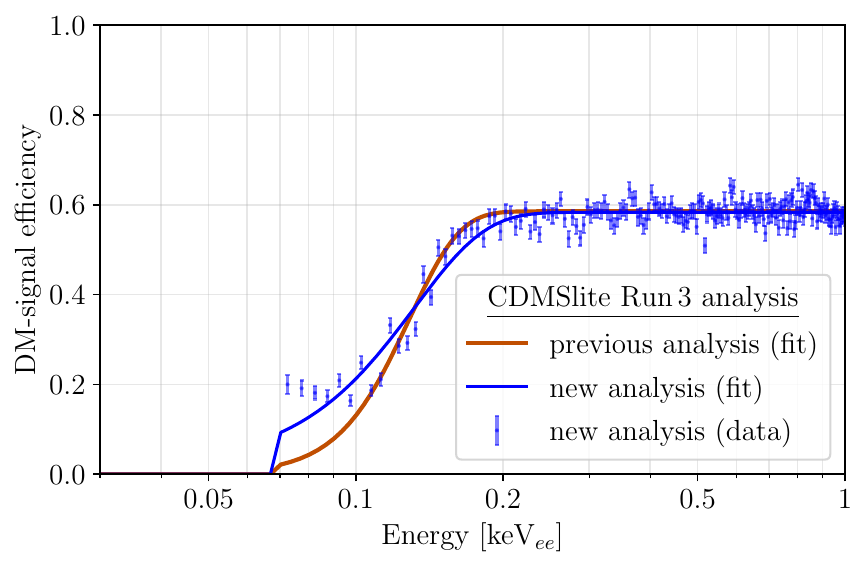}
    \caption{DM-signal efficiency obtained by consecutively applying the trigger efficiency, the quality cut efficiency, and the radial cut efficiency. The previous analysis using the bifurcated LFN cut is shown in orange and the new analysis with the CNN-based LFN cut is shown in blue. The solid lines represent the fit to the data following the parametrization given in Eq.~\ref{eq:final_signal_efficiency}.}
    \label{fig:final_efficiency_old_new}
\end{figure}

Figure~\ref{fig:DQ_efficiency_old_new} shows the comparison of the signal efficiency of the quality cuts for the previous analysis and the new analysis incorporating the CNN-based LFN cut, which exhibits gain in the signal efficiency without compromising the LFN-background rejection.

To obtain the final DM-signal efficiency, the radial cut efficiency must also be included. The radial cut was developed during the  \mbox{CDMSlite} Run\,3 analysis to remove events from the outer detector volume, where field non-uniformities lead to reduced NTL amplification. Figure \ref{fig:radial_efficiency} shows the radial cut efficiency developed for the CDMSlite Run\,3 analysis, where the maximum efficiency is $\sim$60\%, with an analysis threshold of 70\,\evee{}. 
The final DM-signal efficiency using the previous LFN cut and the newly developed cut using the CNN classifier is shown in Fig.~\ref{fig:final_efficiency_old_new}. The efficiencies, parametrized using Eq.~\ref{eq:final_signal_efficiency} with parameters $\mu_1$, $\mu_2$, $\mu_3$ are found to be 0.293$\pm$0.015, 0.122$\pm$0.002, 0.029$\pm$0.002 for the previous analysis, and 0.292$\pm$0.002, 0.116$\pm$0.002, and 0.046$\pm$0.003 for the new analysis.

In both cases, the overall efficiency is reduced due to the radial cut efficiency, which also imposes an analysis threshold of 70\,\evee{}. Despite this, the new analysis achieves an improved signal efficiency up to $\sim$0.12\,\kevee{}. The decrease in signal efficiency between $\sim$0.12\,\kevee{} and 0.26\,\kevee{} in the new analysis compared to the previous analysis results primarily from the choice of the fit \mbox{function}. 

\subsection{DM sensitivities}\label{sec:dm_sensitivities}

The CDMSlite Run\,3 analysis employed a \acrshort{plr} method to calculate exclusion limits on the WIMP-nucleon cross-section \cite{CDMSliter3}. However, in this work, the primary objective is to investigate the impact of the CNN-based LFN cut, which is why a simpler statistical approach such as the \acrshort{oi} method \cite{OI_Yellin} is utilized. 
The region of interest for this analysis is between the analysis threshold of 0.07\,\kevee{} and 2\,\kevee{}. The differences in the event counts from the observed energy spectra directly lead to differences in the sensitivities derived using the \acrshort{oi} method.

\begin{figure}
    \centering
    \includegraphics[width=\linewidth]{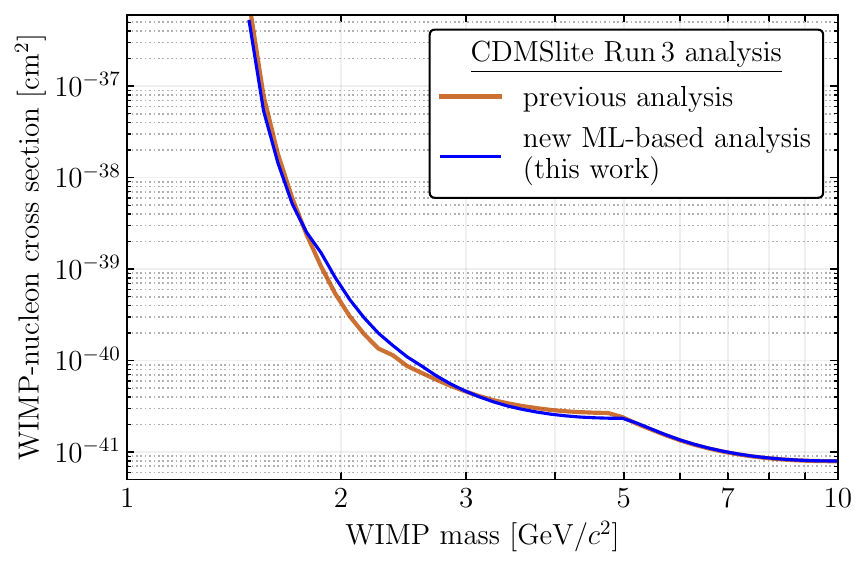}
    \caption{CDMSlite Run\,3 OI-based sensitivities 90\,\% C.L. for the spin-independent WIMP-nucleon cross-section using the previous analysis pipeline with the bifurcated LFN cut (orange) and the new analysis with the CNN-based LFN cut (blue).}
    \label{fig:WIMP_SI_final_limit}
\end{figure}

Figure~\ref{fig:WIMP_SI_final_limit} shows the calculated 90\% Confidence Level (C.L.) sensitivities on the \acrfull{si} WIMP–nucleon cross-section using the \acrshort{oi} method for both the previous analysis using the bifurcated LFN cut and the new analysis using the CNN-based LFN cut. Although the overall profiles are similar, the previous analysis shows better sensitivity in the mass range of 1.8~–~3\,\gevc{}. This is due to the presence of additional events at low energies in the new analysis. The new analysis has slightly better sensitivities between 3 and 5\,\gevc{} due to the improvement in the signal efficiency below $\sim$0.12\,\kevee{} as seen in Fig.~\ref{fig:final_efficiency_old_new}. Overall, the improvement achieved through the implementation of the CNN-based LFN cut does not translate into an overall improvement of the exclusion sensitivity due to the limited efficiency of the radial cut.

 To investigate the impact of the radial cut, case studies involving the removal of the radial cut are presented in the next section.

\section{Case Studies}\label{sec:case_studies}
In this section, three case studies are presented. As discussed previously, the radial cut efficiency significantly reduces the signal efficiency gain achieved by replacing the bifurcated LFN cut from the previous analysis with the CNN-based classifier. Therefore, the first case study explores the impact of removing the radial cut. The second case study investigates the effect of applying a more stringent classification criterion by increasing the signal probability threshold from 0.9 to 0.99. The third case study examines the direct impact of the signal efficiency on the calculated sensitivities by assuming zero observed events, thereby highlighting the effect of the signal efficiency while omitting the influence of purity.

\subsection{Impact of the radial cut}
\begin{figure}[t]
	\centering
	\includegraphics[width=\linewidth]{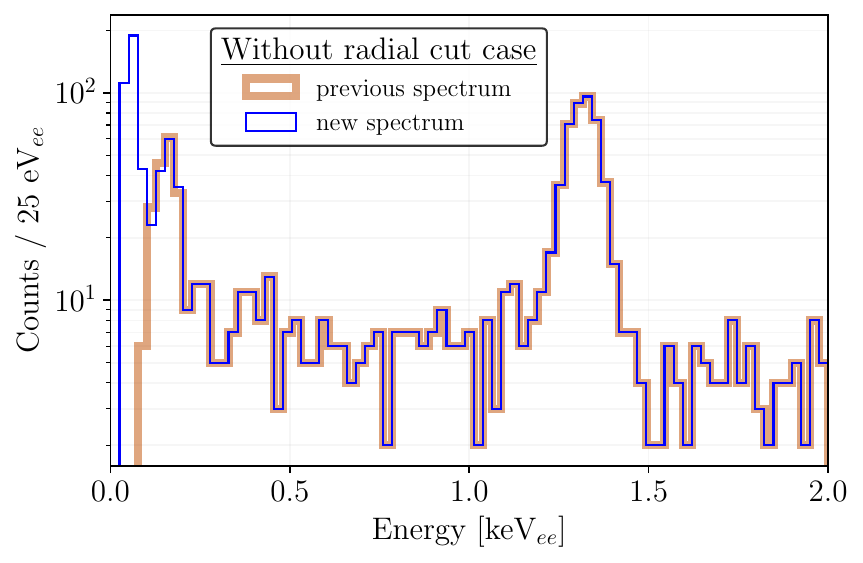}
	\caption{Energy spectrum of CDMSlite Run\,3 using the previous analysis with the bifurcated LFN cut (orange), and the new analysis with CNN-based LFN cut (blue) without the application of the radial cut.}
	\label{fig:final_spectra_woRad}
\end{figure}

This case study investigates the impact of removing the radial cut from the event selection on the DM sensitivity calculation for both the previous analysis and the new CNN-based analysis. The DM-signal efficiencies used here correspond to the quality-cut efficiencies shown in Fig.~\ref{fig:DQ_efficiency_old_new}. The energy spectra without applying the radial cut are shown in Fig.~\ref{fig:final_spectra_woRad}. A higher event rate is observed at 0.05\,\kevee{} in the new analysis, which is not present in the previous analysis. This is partly due to the threshold of the previous analysis being set at $\sim$0.062\,\kevee{} by the bifurcated LFN cut. Additionally, the \acrshort{cnn} was not trained on events below 0.07\,\kevee{}, and therefore caution is required when interpreting events in this region. A closer inspection of these events showed that not all events are signal-like, instead, they include noise events with glitch-like features and pulse-like features in the non-trigger region of the trace which the CNN network was not exposed to during training. As the DM sensitivities are computed using the \acrshort{oi} method, the presence of these additional events in the new analysis leads to weaker constraints.

\begin{figure}
	\centering
	\includegraphics[width=\linewidth]{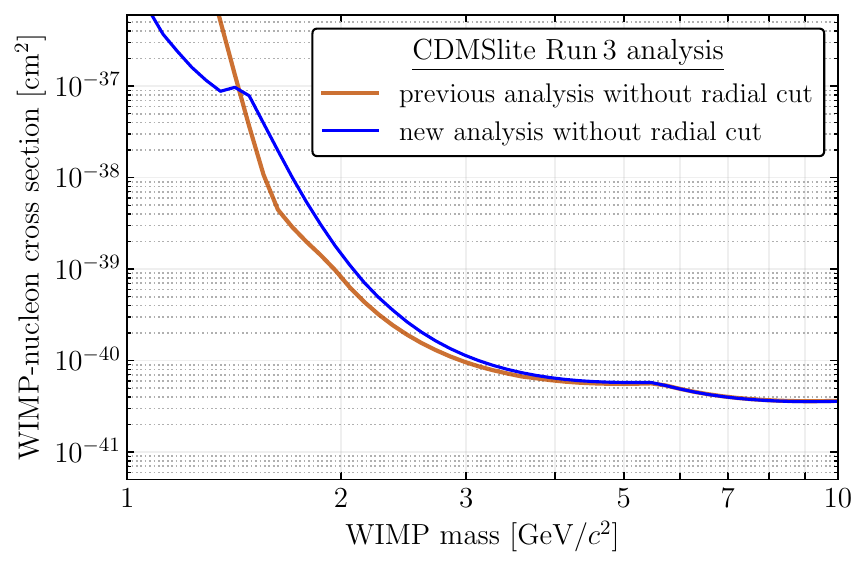}
	\caption{90\,\% C.L. OI-based sensitivities on the spin-independent WIMP-nucleon cross-section for CDMSlite Run\,3 without radial cut using the previous analysis with the bifurcated LFN cut (orange) and the new analysis with the CNN-based LFN cut (blue). }
	\label{fig:final_limit_woRad_85radial}
\end{figure}

In the absence of the radial cut, the effects of reduced NTL amplification must be properly accounted for in the response modeling applied to the expected signal model when calculating the DM sensitivities. Based on the radial cut efficiency study, $\sim$15\% of signal events do not receive full NTL gain and  therefore are not reconstructed at their nominal energies in the energy spectrum \cite{CDMSliter3}. In other words, about $\sim$85\% of events are expected to have full NTL amplification. This fraction ($\sim$85\%) is the NTL efficiency of the detector as described in Sec.~\ref{sec:cdmsliter3_efficiencies}, which quantifies the number of events to be reconstructed at the correct energy. This effect is incorporated by scaling the expected DM-signal model by the NTL efficiency. The corresponding DM sensitivities are shown in Fig.~\ref{fig:final_limit_woRad_85radial}. This scaling of the signal model accounts for the loss of the DM-signal from the nominal energy due to the events with reduced NTL amplification. It is observed that both the sensitivities from the previous and new analysis exclude higher cross-sections for DM masses above 4\,\gevc{} due to the additional events (reduced NTL gain events) in the energy spectrum arising from removing the radial cut. Between DM masses of 1.3\,\gevc{}~–~4\,\gevc{}, the previous analysis excludes a lower cross-section compared to the new analysis due to the absence of events at 0.05\,\kevee{} compared to the new analysis. Below DM masses of 1.3\,\gevc{}, the new analysis outperforms the previous analysis, reaching up to three orders of magnitude improvement in the DM exclusion sensitivities, which was not observed when the radial cut was applied.  This indicates the need for improved radial reconstruction to remove the reduced NTL gain events. The new analysis is also able to probe DM masses as low as 0.8\,\gevc{}, which is lower than the previous analysis and was an important goal of this work.

\subsection{Application of a stringent cut}

The selection of the signal probability threshold plays a crucial role in this analysis. A threshold of 0.5 provides a balanced trade-off between signal efficiency and purity. However, a higher threshold suppresses LFN-background misidentification further and is preferred in this analysis. 

\begin{figure}
	\centering
	\includegraphics[width=\linewidth]{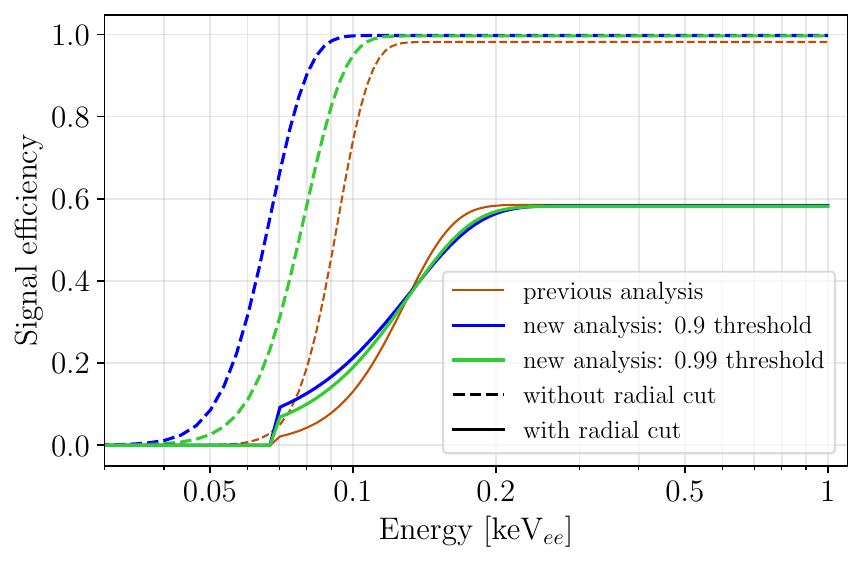}
	\caption{Comparison of the DM-signal efficiencies of \mbox{CDMSlite} Run\,3 with (solid) and without (dashed) the application of the radial cut for the previous analysis using the bifurcated LFN cut (orange), and new analysis using the CNN-based LFN cut with signal probability thresholds of 0.9 (blue) and 0.99 (green). }
	\label{fig:final_efficiency_w0p99_comparison_w_woRad}
\end{figure}

\begin{figure}
	\centering	\includegraphics[width=\linewidth]{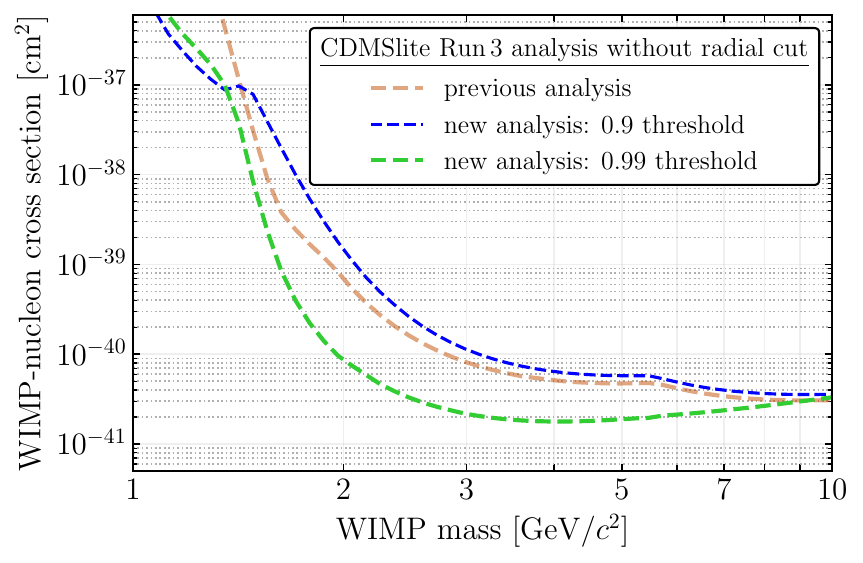}
	\caption{Effect of applying a signal probability threshold of 0.99 and excluding the radial cut on the 90\,\% C.L. OI-based sensitivities on the spin-independent WIMP-nucleon cross-section for CDMSlite Run\,3. Results from the previous analysis with the bifurcated LFN cut (orange) and the new analysis with the CNN-based LFN cut by placing a signal probability threshold at 0.9 (blue) and at 0.99 (green) are shown.}      
	\label{fig:final_limit_worad_w0p99}
\end{figure}

In this case study, a more stringent signal probability threshold of 0.99 is investigated. The DM-signal efficiency with and without applying the radial cut is shown in Fig.~\ref{fig:final_efficiency_w0p99_comparison_w_woRad}. It can be observed that the signal probability threshold of 0.99 has lower signal efficiency than the 0.9 threshold, but still performs better than the previous analysis with the bifurcated LFN cut. With the radial cut applied, the 0.99 threshold produces similar DM sensitivities compared to threshold at 0.9 and the previous analysis.  The DM sensitivities for the 0.99 threshold, together with the 0.9 threshold and the previous analysis are shown in Fig.~\ref{fig:final_limit_worad_w0p99} when the radial cut is not applied. For DM masses below $\sim$1.35\,\gevc{}, the 0.9 threshold performs better than the 0.99 threshold due to the higher signal efficiency at the lowest energies, as seen in Fig.~\ref{fig:final_efficiency_w0p99_comparison_w_woRad}. A significant improvement in sensitivity is observed in the 0.99 threshold case for DM masses between $\sim$1.35\,\gevc{} and 10\,\gevc{}, which results not only from the improved signal efficiency relative to the previous analysis, but also the following important reason: the events that are removed by the stringent LFN cut with a 0.99 signal probability threshold, which previously passed both the bifurcated LFN cut and the 0.9 threshold in the CNN-based approach, have energies between 0.08\,\kevee{} and 0.17\,\kevee{}. As the OI algorithm selects energy intervals in this energy range for the entire mass range probed, the reduced event count directly improves the sensitivity.  

However, it should be noted that the CNN model was not trained for energies below 0.07\,\kevee{}, and results in this energy range should therefore be treated with caution. Overall, this study demonstrates the performance of the CNN-based LFN classifier under a stringent classification threshold and highlights its potential when not limited by the efficiency of the radial cut.

\subsection{In the absence of any events observed}
\begin{figure}
	\centering
	\includegraphics[width=\linewidth]{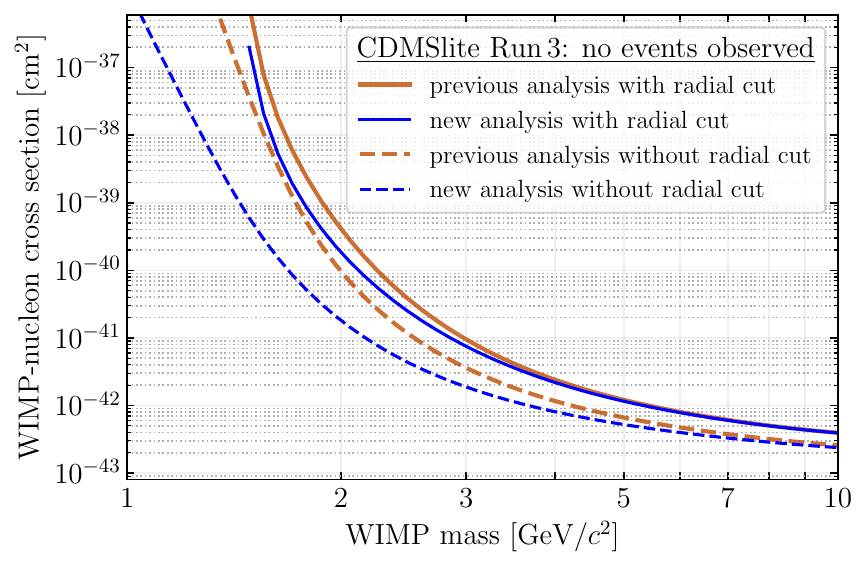}
	\caption{90\,\% C.L. OI-based sensitivities on the spin-independent WIMP-nucleon cross-section for CDMSlite Run\,3 with the assumption of no observed events. The previous analysis using the bifurcated LFN cut and the new analysis with the CNN-based LFN cut are shown in orange and blue, respectively. The solid lines represent the cases where the radial cut is applied, and the dashed lines represent the cases without the radial cut.}
	\label{fig:no_events_wRad_woRad}
\end{figure}

In this case study, the translation of the signal efficiencies to DM sensitivities is explored under the assumption of zero observed events. This is motivated by limitations of the new method, as it was not trained on events below 0.07\,\kevee{}, leading to an increased event count in the low-energy region when the radial cut is not applied as seen in Fig.~\ref{fig:final_spectra_woRad}. Consequently, the full potential of the CNN-based cut cannot be directly assessed in the presence of these additional events. Thus, this study assumes that no events are observed in the energy spectrum. The corresponding DM sensitivities are shown in Fig.~\ref{fig:no_events_wRad_woRad}.\\
With the radial cut applied, there is no improvement in the low-mass DM search reach, as the signal efficiency is limited by the radial-cut energy threshold at 0.07\,\kevee{}. Under these conditions, the previous and new analyses produce comparable exclusion sensitivities, although the new analysis performs slightly better for DM masses below 3.5\,\gevc{}.  In the absence of the radial cut, both analyses show improvement in the low-mass DM search, with the new analysis demonstrating significant improvement in the sensitivity down to a DM mass of 0.8\,\gevc{} and achieving three orders of magnitude improvement in the DM exclusion sensitivities below DM masses of 1\,\gevc{}. This reflects its ability to explore events in the low-energy region. For DM masses of 1\,\gevc{}~–~2\,\gevc{}, the DM sensitivities in the new analysis are improved by up to two orders of magnitude. For intermediate DM masses between 2\,\gevc{}~–~10\,\gevc{}, the new analysis performs slightly better than the previous analysis.

\section{Conclusion}

\acrshort{lfn} events are characterized by variable pulse shapes and high similarity to low-energy signal events. This posed a significant challenge to remove them as over 99\% of the events in the CDMSlite Run\,3 dataset consist of LFN events. Although the previous analysis was able to effectively remove these events, it resulted in a substantially reduced signal efficiency at low energies.

In this work, a \acrshort{cnn}-based classifier was developed to identify and remove LFN events. The multi-channel configuration provided the best performance by achieving 100\% signal purity on the test dataset along with improved signal efficiency compared to the previous analysis. The performance of the new method was evaluated using \acrshort{oi}-based DM sensitivity calculations.

The application of the radial cut was found to limit the potential of the new method by reducing the signal efficiency, which highlights the need for better radial reconstruction to handle events with reduced-NTL gain at larger radii. When the radial cut is removed, allowing events with reduced-NTL gain to remain in the energy spectrum, the CNN-based analysis achieves sensitivity to DM masses down to 0.8\,\gevc{}, which was not accessible in the previous analysis. In addition, for DM masses below 1.3\,\gevc{}, the new analysis showed up to three orders of magnitude improvement in the DM exclusion sensitivities compared to the previous analysis. 

This study also presented a novel method for generating simulated signal traces, which can be used in the future to validate both conventional and alternative simulation techniques. 

As a future application, the CNN-based LFN identification can be extended to remove other types of noise background by retraining the ML model with appropriate datasets.  This work provides a proof of concept using the CDMSlite Run\,3 that is applicable for the upcoming SuperCDMS SNOLAB experiment and future datasets \cite{snolab} \cite{cute}.

\section{Acknowledgments}
The SuperCDMS collaboration gratefully acknowledges technical assistance from the staff of the Soudan Underground Laboratory and the Minnesota Department of Natural Resources. The CDMSlite detectors were fabricated in the Stanford Nanofabrication Facility, which is a member of the National Nanofabrication Infrastructure Network, sponsored and supported by the NSF. Funding and support were received from the National Science Foundation, the U.S. Department of Energy (DOE), Fermilab URA Visiting Scholar Grant No. 15-S-33, NSERC Canada, the Canada First Excellence Research Fund, the Arthur B. McDonald Institute (Canada),  the Department of Atomic Energy Government of India (DAE), the Department of Science and Technology (DST, India) and the DFG (Germany) - Project No. 420484612 and under Germany's Excellence Strategy - EXC 2121 ``Quantum Universe" – 390833306.   Femilab is operated by Fermi Research Alliance, LLC,  SLAC is operated by Stanford University, and PNNL is operated by the Battelle Memorial Institute for the U.S. Department of Energy under contracts DE-AC02-37407CH11359, DE-AC02-76SF00515, and DE-AC05-76RL01830, respectively.

\appendix
\section{CNN model}\label{appendix:cnn}
\raggedbottom
The parameters optimized using a hyperparameter search, which was performed using random search \cite{random_search} and Bayesian optimization \cite{bayesian}, are the number of convolutional layers, the number of filters, kernel size, stride, the number of dense layers, and the use of normalization, regularization, and dropout.  The parameter combination that achieves the highest accuracy with minimal deviation is selected as the final model. The optimized hyperparameters for single-, PT-, and multi-channel are shown in Table.~\ref{tab:hyperparameter_opt_CNN}.
\begin{table}
    \centering
    \caption{Best-performing parameter combinations from the hyperparameter optimization of the CNN architecture for single-, PT-, and multi-channel models. The first two columns list the CNN architecture layers and their corresponding input parameters, respectively, and the last three columns present the optimized parameters for the three models. L2 denotes L2 regularization applied to the bias and kernel regularizers \cite{L2}.}
    \begin{tabular}{c|c|ccc}
        \textbf{Layer}                  & \textbf{Layer parameters}         & \textbf{single}               & \textbf{PT}                   & \textbf{multi}                \\ \hline\hline
        \multirow{4}{*}{Conv1D} & number of filters (x 32) & 1                    & 3                    & 3                    \\
        & kernel size              & 1000                 & 376                  & 200                  \\
        & stride                   & 1                    & 15                   & 18                   \\
        & kernel regularizer (L2)  & 0.01                 & 0.01                 & 0.01                 \\
        & bias regularizer (L2)    & -                    & 0.01                 & 0.01                 \\
        & activation               & relu               & relu               & relu               \\ \hline
        \multirow{4}{*}{Conv1D} & number of filters (x 32) & 2                    & 6                    & 6                    \\
        & kernel size              & 500                  & 188                  & 100                  \\
        & stride                   & 1                    & 1                    & 1                    \\
        & kernel regularizer (L2)  & 0.01                 & 0.01                 & 0.01                 \\
        & bias regularizer (L2)    & -                    & 0.01                 & 0.01                 \\
        & activation               & relu               & relu               & relu               \\ \hline
        GAP                     & -                        & \multicolumn{1}{l}{} & \multicolumn{1}{l}{} & \multicolumn{1}{l}{} \\ \hline
        \multirow{2}{*}{Dense}  & units (x32)              & 2                    & 6                    & 6                    \\
        & kernel regularizer (L2)  & 0.01                 & 0.01                 & 0.01                 \\
        & bias regularizer (L2)    & -                    & 0.01                 & 0.01                 \\
        & activation               & relu               & relu               & relu               \\ \hline
        Dropout                 & rate                     & 0.1                  & 0.276                & 0                    \\ \hline
        \multirow{2}{*}{Dense}  & units (x32)              & 1                    & 3                    & 3                    \\
        & kernel regularizer (L2)  & 0.01                 & 0.01                 & 0.01                 \\
        & bias regularizer (L2)    & -                    & 0.01                 & 0.01                 \\
        & activation               & relu               & relu               & relu               \\ \hline
        Dropout                 & rate                     & 0.1                  & 0.276                & 0                    \\ \hline
        \multirow{2}{*}{Dense}  & units                 & 1                    & 1                    & 1                    \\
        & activation               & sigmoid            & sigmoid            & sigmoid            \\ \hline
    \end{tabular}

    \label{tab:hyperparameter_opt_CNN}
\end{table}

\section{CNN-LSTM model}\label{appendix:cnnlstm}
\raggedbottom
The model parameter search was similar to those of the CNN architecture, with the addition of LSTM layers and batch normalization for training regularization. The optimized hyperparameters for the single-, PT-, and multi-channel configurations are summarized in Table.~\ref{tab:hyperparameter_opt_CNNLSTM}.

\begin{table}
	\centering
    \caption{Best-performing parameter combinations from the hyperparameter optimization of the CNN-LSTM architecture for single-, PT-, and multi-channel models. The first two columns list the CNN-LSTM architecture layers and their corresponding input parameters, respectively, and the last three columns present the optimized parameters for the three models. L2 denotes L2 regularization applied to the bias and kernel regularizers. }
	\begin{tabular}{cc|ccc}
		\multicolumn{1}{c|}{\textbf{Layers}}                  & \textbf{Layer parameters   }      & \textbf{single}    & \textbf{PT}        & \textbf{multi}     \\ \hline\hline
		\multicolumn{1}{c|}{\multirow{4}{*}{Conv1D}} & number of filters (x 32) & 1         & 1         & 2         \\
		\multicolumn{1}{c|}{}                        & kernel size              & 600       & 600       & 200       \\
		\multicolumn{1}{c|}{}                        & stride                   & 20        & 20        & 20        \\
		\multicolumn{1}{c|}{}                         & kernel regularizer (L2)  & -                 & 0.01                 & 0.01                 \\
		\multicolumn{1}{c|}{}                         & bias regularizer (L2)    & -                    & 0.01                 & 0.01                 \\
		\multicolumn{1}{c|}{}                        & activation               & relu    & relu    & relu    \\ \hline
		\multicolumn{2}{c|}{BatchNormalization}                                 & No        & Yes       & Yes       \\ \hline
		\multicolumn{1}{c|}{Dropout}                 & rate                     & 0.1       & 0.1       & 0.2       \\ \hline
		\multicolumn{1}{c|}{\multirow{4}{*}{Conv1D}} & number of filters (x 32) & 2         & 2         & 4         \\
		\multicolumn{1}{c|}{}                        & kernel size              & 300       & 300       & 100       \\
		\multicolumn{1}{c|}{}                        & stride                   & 1         & 1         & 1         \\
		\multicolumn{1}{c|}{}                         & kernel regularizer (L2)  & -                 & 0.01                 & 0.01                 \\
		\multicolumn{1}{c|}{}                         & bias regularizer (L2)    & -                    & 0.01                 & 0.01                 \\
		\multicolumn{1}{c|}{}                        & activation               & relu    & relu    & relu    \\ \hline
		\multicolumn{2}{c|}{BatchNormalization}                                 & No        & Yes       & Yes       \\ \hline
		\multicolumn{1}{c|}{Dropout}                 & rate                     & 0.1       & 0.1       & 0.2       \\ \hline
		\multicolumn{1}{c|}{\multirow{2}{*}{LSTM}}   & units (x32)              & 2         & 2         & 4         \\
		\multicolumn{1}{c|}{}                        & return sequence          & True       & True       & True       \\ \hline
		\multicolumn{2}{c|}{BatchNormalization}                                 & No        & Yes       & Yes       \\ \hline
		\multicolumn{1}{c|}{Dropout}                 & rate                     & 0.2       & 0.2       & 0.1       \\ \hline
		\multicolumn{1}{c|}{\multirow{2}{*}{LSTM}}   & units (x32)              & 2         & 2         & 4         \\
		\multicolumn{1}{c|}{}                        & return sequence          & True       & True       & True       \\ \hline
		\multicolumn{2}{c|}{BatchNormalization}                                 & No        & Yes       & Yes       \\ \hline
		\multicolumn{1}{c|}{Dropout}                 & rate                     & 0.2       & 0.2       & 0.3       \\ \hline
		\multicolumn{1}{c|}{\multirow{2}{*}{Dense}}  & units (x32)              & 2         & 2         & 4         \\
		\multicolumn{1}{c|}{}                         & kernel regularizer (L2)  & -                 & 0.01                 & 0.01                 \\
		\multicolumn{1}{c|}{}                         & bias regularizer (L2)    & -                    & 0.01                 & 0.01                 \\
		\multicolumn{1}{c|}{}                        & activation               & relu    & relu    & relu    \\ \hline
		\multicolumn{1}{c|}{Dropout}                 & rate                     & 0.3       & 0.3       & 0         \\ \hline
		\multicolumn{1}{c|}{\multirow{2}{*}{Dense}}  & units                    & 1         & 1         & 1         \\
		\multicolumn{1}{c|}{}                        & activation               & sigmoid & sigmoid & sigmoid \\ \hline
	\end{tabular}
	\label{tab:hyperparameter_opt_CNNLSTM}
\end{table}
\newpage
\bibliographystyle{aps}
\bibliography{bibliography}

\end{document}